\documentclass[lettersize,journal]{IEEEtran}

\usepackage{amsmath,amsfonts}
\usepackage{algorithmic}
\usepackage{array}
\usepackage[caption=false,font=footnotesize,labelfont=rm,textfont=rm]{subfig}
\usepackage{textcomp}
\usepackage{stfloats}
\usepackage{url}
\usepackage{verbatim}
\usepackage{graphicx}
\def\BibTeX{{\rm B\kern-.05em{\sc i\kern-.025em b}\kern-.08em
    T\kern-.1667em\lower.7ex\hbox{E}\kern-.125emX}}
\usepackage{balance}

\usepackage{amssymb} 
\usepackage{booktabs}
\usepackage{upgreek}
\usepackage{cite}
\usepackage{setspace}

\begin{document}
\title{Waveform-Domain Adaptive Matched Filtering for Suppressing Interrupted-Sampling Repeater Jamming}
\author{Hanning Su, Qinglong Bao, Jiameng Pan, Fucheng Guo, and Weidong Hu
\thanks{This work was supported by the National Science Foundation of China under Grant 62231026. (Corresponding author: Qinglong Bao.)

Hanning Su, Qinglong Bao, Jiameng Pan, Fucheng Guo and Weidong Hu are with the College of Electronic Science and Technology, National University of Defense Technology, Changsha 410073, China (e-mail:hanningsu18@yahoo.com; baoqinglong@nudt.edu.cn; panjiameng@nudt.edu.cn; gfcly@21cn.com; wdhu@nudt.edu.cn;)}}

\markboth{Journal of \LaTeX\ Class Files,~Vol.~18, No.~9, September~2020}%
{How to Use the IEEEtran \LaTeX \ Templates}

\maketitle

\begin{abstract}
The inadequate adaptability to flexible interference scenarios remains an unresolved challenge in the majority of techniques utilized for mitigating interrupted-sampling repeater jamming (ISRJ). Matched filtering system based methods is desirable to incorporate anti-ISRJ measures based on prior ISRJ modeling, either preceding or succeeding the matched filtering. Due to the partial matching nature of ISRJ, its characteristics are revealed during the process of matched filtering. Therefore, this paper introduces an extended domain called the waveform domain within the matched filtering process. On this domain, an adaptive matched filtering model, known as the waveform-domain adaptive matched filtering (WD-AMF), is established to tackle the problem of ISRJ suppression without relying on a pre-existing ISRJ model. The output of the WD-AMF encompasses an adaptive filtering term and a compensation term. The adaptive filtering term encompasses the adaptive integration outcomes in the waveform domain, which are determined by an adaptive weighted function. This function, akin to a collection of bandpass filters, decomposes the integrated function into multiple components, some of which contain interference while others do not. The compensation term adheres to an integrated guideline for discerning the presence of signal components or noise within the integrated function. The integration results are then concatenated to reconstruct a compensated matched filter signal output. Simulations are conducted to showcase the exceptional capability of the proposed method in suppressing ISRJ in diverse interference scenarios, even in the absence of a pre-existing ISRJ model.
\end{abstract}


\section{INTRODUCTION}
The Interrupted-Sampling Repeater Jamming (ISRJ) represents a form of intra-pulse interference, wherein the jammer samples a brief segment of the radar waveform and promptly retransmits it \cite{1,2}. The jamming signals exhibit strong coherence with the actual target echo, resulting in the appearance of both genuine and spurious target peaks in the range profile obtained through matched filtering. By employing flexible jamming parameters, the jammer has the capability to generate a variable number of false targets with varying amplitude and positions \cite{3,4,5,6}. 

One prominent focus in the research on ISRJ suppression is the enhancement of interference suppression signal-to-noise ratio (SNR) while minimizing the loss of target SNR, thereby improving adaptability to challenging scenarios characterized by limited snapshots, low SNR, flexible signal-to-jamming ratio (SJR), and varying ISRJ modulation schemes. Several approaches have been proposed to address these requirements, such as orthogonal waveforms and filter design methods \cite{7,8,9,10}, as well as time-frequency domain filtering techniques \cite{11,12,13,14}. A common characteristic among these methods is their reliance on time-domain matched filtering systems, which establish mappings from the time-domain signal to the compressed pulse signal and assume reversibility of these mappings. Exploiting this assumption, the filter's inputs or outputs can be matched with pre-established mappings to mitigate ISRJ and achieve a range profile devoid of interference. Different methods emerge from distinct matching criteria, such as the utilization of a separable convex optimization scheme for joint waveform and mismatched filter design \cite{7,8,9,10}, or the implementation of band-pass filtering based on time-frequency analysis for time-frequency domain filtering methods \cite{11,12,13,14}. The effectiveness of these methods relies significantly on prior information pertaining to the ISRJ model, encompassing the jammer's operational mode, modulation scheme, and operational parameters.

The time-domain matched filtering system can experience various imperfections attributed to the indirect modulation characteristics of ISRJ. Furthermore, ISRJs can display diverse patterns in different durations, reflecting their specific objectives \cite{3,4,5,6,15}. Consequently, the preformulated mappings of inputs and outputs of the filter in practical systems become considerably complex in ISRJ suppression methods. Some of these imperfections are too complicated to be modeled accurately, and the inaccurate modeling may pose significant negative influence on the performance of ISRJ suppression. To facilitate the implementation of the methods, cognitive models are developed to depict the operational characteristics of flexible interference scenarios \cite{16,17,18,19}, while deconvolution processes are employed to estimate the crucial parameters of ISRJ \cite{20}. 

Indeed, cognitive model-based methods are limited in adaptability due to their reliance on the accuracy of the cognitive model, which restricts their potential for broader application. In reference \cite{21}, a neural network is introduced to extract segments of the signal free from jamming interference, enabling the generation of a band-pass filter. This adaptive approach circumvents the need for prior information about the interference. However, it is necessary to further validate the performance of the network using real radar measurements, as it has currently been trained only with simulated data. Additionally, the band-pass filter method is specifically applicable to stretched echoes of linear frequency modulated (LFM) waveforms, necessitating further investigation to address ISRJ suppression in the presence of complex waveforms.

In recent research \cite{22}, an integration decomposition method is introduced to tackle the challenge of recognizing false targets caused by ISRJ. It establishes an intrinsic integration sequence starting from the received echo and derives a nonlinear mapping to an antiderivative of an energy function. This derived mapping is subsequently employed to extract the characteristics of ISRJ. This suggests the existence of potential adaptive discriminative features between ISRJ and the actual echo signal within the micro-domain of the matched filtering system. However, in reference \cite{22}, only preliminary pattern classification based on these discriminative features is conducted, without delving into a comprehensive discussion of the mathematical principles underlying this phenomenon. As a result, further investigation into this potential micro-domain remains necessary.

ISRJ fully exploits the deficiencies exhibited by the accumulated output of the matched filtering process. Moreover, from the perspective of the data structure involved in the convolution process of matched filtering, the ISRJ, being only partially matched, manifests its characteristics within the data structure. Hence, we define the waveform domain as the domain of the process data and establish a new matched filtering model upon it. In this manuscript, we present a comprehensive method called Waveform-Domain Adaptive Matching Filtering (WD-AMF) as a solution to the broader problem of ISRJ suppression. In WD-AMF, our focus shifts from the macroscopic input and output waveforms of the time-domain matched filtering to the integrated function within the convolution operation in the waveform domain. To effectively suppress ISRJ while preserving the output gain of the echo signal, we employ a robust adaptive algorithm in the waveform domain. 

The remaining sections of this paper are organized into six parts. In Section \uppercase\expandafter{\romannumeral2}, we establish the problem formulation for mitigating ISRJ. Section \uppercase\expandafter{\romannumeral3} introduces the framework of the WD-AMF, encompassing the relevant definitions and representations of intermediate variables. Section \uppercase\expandafter{\romannumeral4} presents the expressions and characteristics of cumulative waveform coherence functions for the echo signal, ISRJ, and received signal of the LFM waveform. Building upon these findings, a statistical model is formulated to effectively suppress ISRJ. Section \uppercase\expandafter{\romannumeral5} outlines the application of the IMM-KF technique to solve the anti-ISRJ model, providing a comprehensive expression of the WD-AMF. In Section \uppercase\expandafter{\romannumeral6}, simulations are conducted to demonstrate the superior performance of the proposed method in mitigating ISRJ. Finally, in Section \uppercase\expandafter{\romannumeral7}, we conclude this paper.

\section{Problem Formulation}
\subsection{ISRJ model}
The ISRJ signal can be expressed as the product of the transmitted waveform $s(t)$ and the interrupted-sampling function $g(t)$, denoted as $\jmath(t) = g(t) s(t)$. By utilizing the ambiguity function, the output of the matched filter for the ISRJ signal can be represented by the following equation \cite{2}:
\begin{equation}
\jmath_{o}(t)=\sum_{n=-\infty}^{+\infty} f_J T_\jmath \operatorname{Sa}\left(n \pi f_J T_{\jmath}\right) \mathcal{X}\left(t,-n f_J\right)
\label{eq1}
\end{equation}
where $\operatorname{Sa}(x)=\frac{\sin(x)}{x}$, $T_\jmath$ represents the duration of the jamming slice. The interrupted-sampling frequency is denoted by $f_{J}=\frac{1}{T_{J}}$, and $T_{J}$ corresponds to the interrupted-sampling repeater period. The function $\mathcal{X}(t, f_d) = s(t+\tau)s^{\ast}(\tau)e^{j2\pi f_d \tau}\mathrm{d}\tau$ represents the ambiguity function of the transmitted waveform. It is worth noting that for self-defensive repeater interference, due to the utilization of a time-sharing transmit-receive antenna by the jammer, the condition $f_{J}T_\jmath \leqslant 0.5$ holds. 

\subsection{anti-ISRJ model}
Let's assume that a monostatic pulsed Doppler radar transmits a pulse compression waveform $s(t)$. When the self-defensive jammer transmits a jamming signal, the received signal can be expressed as:
\begin{equation}
\begin{aligned}
x(t) =A_ss(t-\tau_s)+A_\jmath\jmath(t-\tau_\jmath)
\end{aligned}
\label{eq2}
\end{equation}
Here, $A_s$ represents the amplitude of the target echo signal, $\tau_s$ denotes the propagation delay of the target. On the other hand, $A_\jmath$ represents the amplitude of the jamming signal, and $\tau_\jmath$ denotes the delay of the interference signal. 

When $x(t)$ traverses a generalized matched filter $h(t)$, characterized by the system function $\mathrm{H}[\cdot]$, within the framework of ISRJ suppression methods based on matched filtering systems, a predefined mapping is employed for both $x(t)$ and $h(t)$. Consequently, an output is generated, which can be expressed as follows:
\begin{equation}
\begin{aligned} z_o(t) & =\mathrm{KH}\{\mathrm{G}[x(t)]\} 
\\ & = \mathrm{G}[x(t)] \otimes \mathrm{K}[h(t)] 
\\ & = \mathrm{G}[s(t-\tau_s)] \otimes \mathrm{K}[h(t)]+\mathrm{G}[\jmath(t-\tau_\jmath)] \otimes \mathrm{K}[h(t)] 
\\ & =\mathrm{KH}\{G[s(t-\tau_s)]\}+\varrho(t)
\end{aligned}
\label{eq4}
\end{equation}
Here, $\mathrm{KH}[\cdot]$ symbolizes the mapping resulting from the application of $\mathrm{K}[\cdot]$ to the system function $\mathrm{H}[\cdot]$. Furthermore, $\mathrm{G}[\cdot]$ represents the mapping in relation to $x(t)$. Additionally, $\varrho(t)$ can be defined as the convolution of $\mathrm{G}[\jmath(t-\tau_\jmath)]$ and $\mathrm{K}[h(t)]$. Hence, an effective strategy to counteract the interference from ISRJ involves maximizing the amplitude of the main lobe in $\mathrm{KH}\{G[s(t-\tau_s)]\}$, while simultaneously minimizing the amplitude of the side lobes in $\mathrm{KH}\{G[s(t-\tau_s)]\}$ and inhibiting the peak amplitude of $\varrho(t)$. 

\section{Waveform-Domain Adaptive Matched Filtering}

Signal time-domain matched filtering is defined as:
\begin{equation}
\begin{aligned}
x_o(t) = \int_{-\infty}^{\infty} x(t-\mu) h(\mu) d \mu
\end{aligned}
\label{eq5}
\end{equation}

\begin{figure}[htbp]
\includegraphics[width=8 cm]{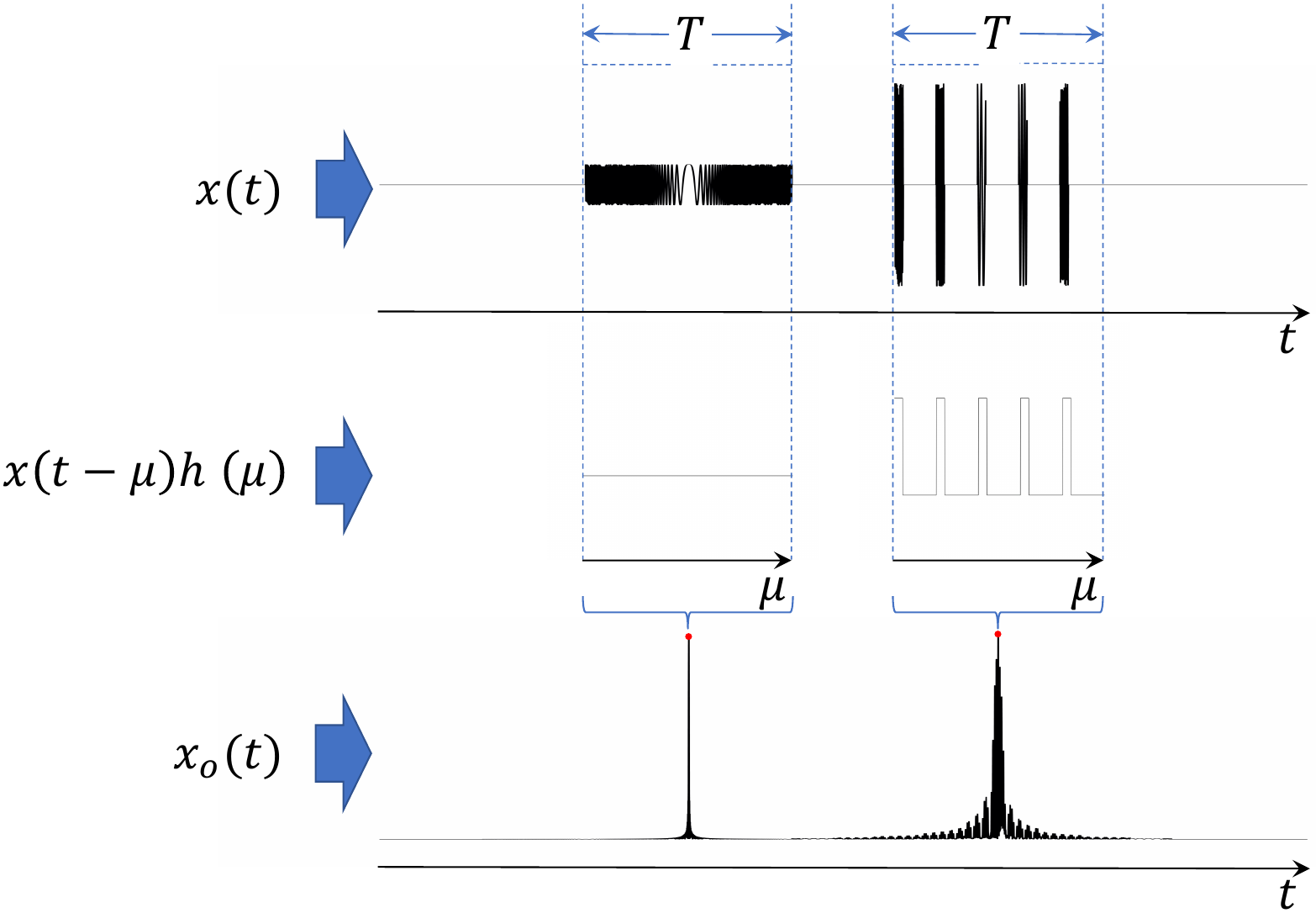}
\centering
\caption{Illustration of the data structure entailed in the convolution process of matched filtering, exemplified using a LFM waveform.
\label{fig0}}
\end{figure}

If $x(\mu)$ and $h(\mu)$ represent signals with finite duration $T$, (\ref{eq5}) can be interpreted as the overall integral of the product of $x(t-\mu)$ and $h(\mu)$ across the fast time variable $\mu$ within the intra-pulse period of $h(\mu)$. By considering (\ref{eq1}), it can be observed that ISRJ fully exploits the deficiencies exhibited by the accumulated output of the matched filtering process. Even if ISRJ is temporally discontinuous, its output after passing through the matched filter can still be represented in the same form as the target echo. However, from the perspective of the data structure involved in the convolution process of matched filtering, the characteristics of ISRJ, being only partially matched, are manifested within the data structure as depicted in Fig. \ref{fig0}.

Consequently, we designate the fast time domain $\mu$ within $h(\mu)$ as the waveform domain, and the function being integrated as the waveform response function (WRF). This term signifies the response that occurs once the waveform traverses the filter and is defined as follows:
\begin{equation}
\begin{aligned}
\upsilon^{(t)}(\mu) &=  x(t-\mu) h(\mu)
\\
&=\upsilon_s^{(t-\tau_s)}(\mu)+\upsilon_\jmath^{(t-\tau_\jmath)}(\mu)
\end{aligned}
\label{eq6}
\end{equation}
where $\mu \in \left[-\frac{T}{2},\frac{T}{2}\right]$, $\upsilon_s^{(t)}(\mu) = A_ss(t-\mu)h(\mu)$ denotes the waveform response function of $s(t)$, and $\upsilon_\jmath^{(t)}(\mu) = A_\jmath\jmath(t-\mu)h(\mu)$ denotes the waveform response function of $\jmath(t)$. Next, our point of interest lies in the instantaneous analytical expression of $\upsilon^{(t)}(\mu)$ in the waveform domain. Divergent from temporal matched filtering of signals, we define the cumulative waveform coherence function (CWCF) as the variable upper limit integration of $\upsilon^{(t)}(\mu)$, in the waveform domain:
\begin{equation}
\begin{aligned}
y^{(t)}(\rho) 
&= \int_{-\infty}^{\rho} \upsilon^{(t)}(\mu) \mathrm{d} \mu
\\
& = y_s^{(t-\tau_s)}(\rho) + y_\jmath^{(t-\tau_\jmath)}(\rho)
\label{eq7}
\end{aligned}
\end{equation}
where $y_s^{(t)}(\rho) = \int_{-\infty}^{\rho} \upsilon_s^{(t)}(\mu) \mathrm{d} \mu$ denotes the CWCF of $\upsilon_s^{(t)}(\mu)$, and $y_\jmath^{(t)}(\rho) = \int_{-\infty}^{\rho} \upsilon_\jmath^{(t)}(\mu) \mathrm{d} \mu$ denotes the CWCF of $\upsilon_\jmath^{(t)}(\mu)$. By comparing (\ref{eq5}) and (\ref{eq7}), it becomes apparent that (\ref{eq7}) is not restricted to a single matched filter value in the time domain, but rather encompasses the integration variable of the entire waveform domain. In the scenario where $\rho\to\infty$, it follows that $y^{(t)}(\rho)=x_o(t)$.

If we define $a^{(t)}(\rho) = \mathrm{Y}\left[y^{(t)}(\rho)\right]$, whcih denotes the weight function of $v^{(t)}$, while $\varsigma^{(t)} = \mathrm{V}\left[y^{(t)}(\rho)\right]$ denotes the compensation term. Then, the WD-AMF can be defined as:
\begin{equation}
z_o(t) = \int_{-\infty}^{\infty} a^{(t)}(\mu)v^{(t)}(\mu) d \mu + \varsigma^{(t)}
\label{eq8}
\end{equation}

The WD-AMF method can be considered as a micro-operation conducted on the conventional matched filter. By adaptively adjusting $a^{(t)}(\mu)$ and $\delta^{(t)}$, the output $z_{o}(t)$ achieves efficient suppression of interference signals while maintaining the original signal energy in an adaptive fashion. 

\section{Cumulative waveform coherence function}
In the subsequent sections, we will deduce and thoroughly examine the analytical expressions for the CWCF of $s(t)$, $\jmath(t)$, and $x(t)$.
\subsection{$y_{s}^{(t)}(\rho)$}
In the context of pulsed Doppler radar systems, assuming that the transmitted waveform is a LFM signal. Consequently, the baseband signal can be mathematically expressed as follows:
\begin{equation}
\begin{aligned}
s(t)=\operatorname{rect}\left(\frac{t}{T}\right) e^{j \pi k t^{2}}
\end{aligned}
\label{eq9}
\end{equation}
Here, the function $\operatorname{rect}(\cdot)$ represents the rectangular window function, $T$ corresponds to the pulse width here, and $k$ corresponds to the chirp rate. The impulse response of the matched filter can be expressed as follows:
\begin{equation}
h(t)=\operatorname{rect}\left(\frac{t}{T}\right) e^{-j \pi k t^{2}}
\label{eq10}
\end{equation}

Combining (\ref{eq7}), we can obtain the expression for CWCF of $s(t)$:
\begin{equation}
\begin{array}{l}
y_s^{(t)}(\rho) 
\\
=\int_{-\infty}^{\rho} s(t-\mu) h(\mu) \mathrm{d} \mu  
\\ 
=\left\{\begin{array}{cl}
c^{(t)}(\rho), 
& \text { when } \alpha_s^{(t)} \leqslant \rho \leqslant \beta_s^{(t)} \\
c_{0}^{(t)}, 
& \text { when } \rho>\beta_s^{(t)} 
\\
0, 
& \text { when } \rho<\alpha_s^{(t)}
\end{array}\right.\end{array}
\label{eq11}
\end{equation}
Here, $\alpha_s^{(t)}=\operatorname{max}\left\{-\frac{T}{2},-\frac{T}{2}+t\right\}$ and $\beta_s^{(t)}=\operatorname{min}\left\{\frac{T}{2},\frac{T}{2}+t\right\}$. $c^{(t)}(\rho)$ and $c_{0}^{(t)}$ are defined as follows:
\begin{subequations}
\begin{equation}
\begin{aligned}
c^{(t)}(\rho) = c_{1}^{(t)}(\rho) \cdot \operatorname{Sa}\left[\pi k t\left(\rho-\alpha_s^{(t)}\right)\right]
\end{aligned}
\label{eq12a}
\end{equation} 
\begin{equation}
\begin{aligned}
c_{1}^{(t)}(\rho) = \left(\rho-\alpha_s^{(t)}\right)\cdot e^{j\pi k t\left(t-\rho-\alpha_s^{(t)}\right)}
\end{aligned}
\label{eq12b}
\end{equation}
\begin{equation}
\begin{aligned}
c_0^{(t)} = c^{(t)}(\beta_{s}^{(t)}) 
\end{aligned}
\label{eq12c}
\end{equation}
\label{eq12}  
\end{subequations}

The derivation of ($\ref{eq11}$) and ($\ref{eq12}$) can be found in ($\ref{Aeq2}$) to ($\ref{Aeq8}$). By focusing solely on the magnitudes of $y_{s}^{(t)}(\rho)$, it becomes evident that for $t\ne 0$ and $\alpha_s^{(t)} \leqslant \rho \leqslant \beta_s^{(t)}$,
\begin{equation}
\left|y_s^{(t)}(\rho)\right| = \left|\frac{\sin\left[\pi kt\left(\rho-\alpha_s^{(t)}\right)\right]}{\pi kt}\right|
\label{eq13}
\end{equation}
which represents a half-wave rectified function. The magnitude is $\left|\frac{1}{\pi kt}\right|<\left|\beta_s^{(t)}-\alpha_s^{(t)}\right|=T$, and it decreases as $\left|t\right|$ increases. Based on the properties of the $\mathrm{Sa}(\cdot)$ function, it can be inferred that when $\rho \rightarrow \infty$, $\left|y_s^{(t)}(\rho)\right|\rightarrow 0$.

In particular, when $t= 0$, $\left|y_s^{(0)}(\rho)\right|$ takes the form of a first-order linear function:
\begin{equation}
\left|y_s^{(0)}(\rho)\right| =y_s^{(0)}(\rho)  = \rho+\frac{T}{2}
\label{eq14}
\end{equation}
where the minimum value of $\left|y_s^{(0)}(\rho)\right|$ is 0, and the maximum value is $T$.

\subsection{$y_{\jmath}^{(t)}(\rho)$}
For ISRJ, the interference signal can be regarded as:
\begin{equation}
\begin{aligned}
\jmath(t)
&=\sum_{n = -\infty}^{\infty} \jmath_n (t)
\\
&=\sum_{n = -\infty}^{\infty}\operatorname{rect}\left(\frac{t-nT_J}{T_\jmath}\right) \cdot  s(t)
\end{aligned}
\label{eq15}
\end{equation}
Here, $\jmath_n(t)$ represents the $n$-th interference slice. Then, CWCF of of $\jmath(t)$ can be expressed as:
\begin{equation}
y_{\jmath}^{(t)}(\rho) = \sum_{n = -\infty}^{\infty} y_{\jmath_{n}}^{(t)}(\rho)
\label{eq16} 
\end{equation}
where $y_{\jmath_{n}}^{(t)}(\rho)$ represents the CWCF of the $n$-th individual interference slice:
\begin{equation}
\begin{array}{l}
y_{\jmath_{n}}^{(t)}(\rho)  
\\ 
=\left\{\begin{array}{cl}
b_n^{(t)}(\rho), 
& \text { when } \alpha_{\jmath_n}^{(t)} \leqslant \rho \leqslant \beta_{\jmath_n}^{(t)} \\ 
b_{n_0}^{(t)}, 
& \text { when } \rho>\beta_{\jmath_n}^{(t)} 
\\ 
0, & \text { when } \rho<\alpha_{\jmath_n}^{(t)}
\end{array}\right.\end{array}
\label{eq17}
\end{equation}
Here, $\alpha_{\jmath_n}^{(t)}=\max\left\{-\frac{T}{2},-\frac{T}{2}+t,-\frac{T_\jmath}{2}+ nT_{J}+t\right\}$ and $\beta_{\jmath_n}^{(t)}=\min\left\{\frac{T}{2},\frac{T}{2}+t,\frac{T_\jmath}{2}+nT_{J}+t\right\}$. $b_{n}^{(t)}(\rho)$ and $b_{n_0}^{(t)}$ are defined as follows:
\begin{subequations}
\begin{equation}
\begin{aligned}
b_n^{(t)}(\rho) = b_{n_1}^{(t)}(\rho) \cdot \operatorname{Sa}\left[\pi k t\left(\rho-\alpha_{\jmath_n}^{(t)}\right)\right]
\end{aligned}
\label{eq18a}
\end{equation} 
\begin{equation}
\begin{aligned}
b_{n_1}^{(t)}(\rho) = \left(\rho-\alpha_{\jmath_n}^{(t)}\right)\cdot e^{j\pi k t\left(t-\rho-\alpha_{\jmath_n}^{(t)}\right)}
\end{aligned}
\label{eq18b}
\end{equation}
\begin{equation}
\begin{aligned}
b_{n_0}^{(t)} = b_n^{(t)}(\beta_{\jmath_n}^{(t)}) 
\end{aligned}
\label{eq18c}
\end{equation}
\label{eq18}  
\end{subequations}

The derivation of ($\ref{eq17}$) and ($\ref{eq18}$) can be found in ($\ref{Beq2}$) to ($\ref{Beq16}$). It is straightforward to infer that when $\beta_{\jmath_{m-1}}^{(t)}<\rho\leqslant\beta_{\jmath_{m}}^{(t)}$, the following equation holds true:
\begin{equation}
\begin{array}{l}
y_{\jmath}^{(t)}(\rho)  
\\ 
=\left\{\begin{array}{cl}
\sum\limits_{n = -\infty}^{m-1} b_{n_0}^{(t)},
& \text { when } \beta_{\jmath_{m-1}}^{(t)} \leqslant \rho < \alpha_{\jmath_m}^{(t)} \\ 
\sum\limits_{n = -\infty}^{m-1} b_{n_0}^{(t)} + b_{m}^{(t)}, 
& \text { when } \alpha_{\jmath_m}^{(t)} \leqslant \rho \leqslant \beta_{\jmath_m}^{(t)}
\end{array}\right.\end{array}
\label{eq19}
\end{equation}

Hence, $y_\jmath^{(t)}(\rho)$ can be expressed as a piecewise function. Specifically, when $t = 0$, we have:
\begin{equation}
\begin{aligned}
&\left|y_{\jmath}^{(0)}(\rho)\right| = y_{\jmath}^{(0)}(\rho)
\\
&=\begin{array}{l}  
\left\{\begin{array}{l}
\sum\limits_{n=-\infty}^{m-1} T_\jmath, \quad \text { when } \beta_{\jmath_{m-1}}^{(0)} \leqslant \rho<\alpha_{\jmath_{m}}^{(0)} 
\\
\rho+\frac{T_\jmath}{2}-mT_J+\sum\limits_{n=-\infty}^{m-1}T_\jmath, \quad \text { when } \alpha_{\jmath_{m}}^{(0)} \leqslant \rho \leqslant \beta_{\jmath_{m}}^{(0)}\end{array}\right.\end{array}
\end{aligned}
\label{eq20}
\end{equation}
which is a stepped function. The minimum value of $\left|y_\jmath^{(0)}(\rho)\right|$ is 0, and the maximum value is $\frac{T_\jmath}{T_{J}}\cdot T$. 

When $t\ne 0$, the expression for $\left|b_n^{(t)}(\rho)\right|$ is given as follows:
\begin{equation}
\left|b_n^{(t)}(\rho)\right| = \left|\frac{\sin \left[\pi kt\left(\rho-\alpha_{\jmath_n}^{(t)}\right)\right]}{\pi kt}\right|
\label{eq21}
\end{equation}
which has a period of $T_{b}^{(t)} = \left|\frac{1}{kt}\right|$. In particular, when $t = t_{\jmath_n} = -\frac{nf_{J}}{k}$, we have:
\begin{equation}
\left|b_n^{(t_{\jmath_n})}(\rho)\right| = \left|\frac{\operatorname{sin}\left[\pi nf_{J}\left(\rho-\alpha_{\jmath_n}^{(t_{\jmath_n})}\right)\right]}{\pi nf_{J}}\right|
\label{eq22}
\end{equation}
and $T_{b}^{(t_{\jmath_n})} = \frac{1}{nf_{J}} = \frac{T_{J}}{n}$. If $T_\jmath \leqslant \frac{T_{J}}{2n}$, then $\left|b_n^{(t_{\jmath_n})}(\rho)\right|$ is a monotonically increasing function. Therefore, $|y_{\jmath}^{(t_{\jmath_n})}(\rho)|$ can also be thought of as a stepped function similar to $|y_{\jmath}^{(0)}(\rho)|$:
\begin{equation}
\begin{aligned}
&\left|y_{\jmath}^{(t_{\jmath_n})}(\rho)\right|
\\
&\approx\begin{array}{l}  
\left\{\begin{array}{l}
\left|\sum\limits_{n=-\infty}^{m-1} b_{n_0}^{(t)}\right|, 
\\
\quad \text { when } \beta_{\jmath_{m-1}}^{(t_{\jmath_n})} \leqslant \rho<\alpha_{\jmath_{m}}^{(t_{\jmath_n})} 
\\
Q\rho+\frac{T_\jmath}{2}-mT_J+\left|\sum\limits_{n=-\infty}^{m-1} b_{n_0}^{(t)}\right|, 
\\
\quad \text { when } \alpha_{\jmath_{m}}^{(t_{\jmath_n})} \leqslant \rho \leqslant \beta_{\jmath_{m}}^{(t_{\jmath_n})}
\end{array}\right.\end{array}
\end{aligned}
\label{eq23}
\end{equation}
where $Q<1$ arises from the fact that the matched filter is the maximum output SNR filter, and the minimum value of $\left|y_s^{(t_{\jmath_n})}(\rho)\right|$ is 0, while the maximum value is $f_{J} T_{\jmath} \operatorname{Sa}\left(n \pi f_{J} T_{\jmath}\right) \mathcal{X}\left(t_{\jmath_n},-n f_{J}\right)$.

When $t \ne t_{\jmath_n}$, the effective accumulation of $|b_n^{(t)}|$ is lacking, resulting in $|y_{\jmath}^{(t)}(\rho)|$ being represented as a piecewise envelope with a smaller maximum amplitude. As the difference $\left|t-t_\jmath\right|$ increases, $|b_{n_0}^{(t)}|$ becomes smaller, and $|y_{\jmath}^{(t)}(\rho)|$ becomes closer to $|y_{s}^{(t)}(\rho)|$, which can be approximated as a stable half-wave rectification function:
\begin{equation}
\begin{array}{l}
\left|y_{\jmath}^{(t)}(\rho)\right|  
\\ 
\approx \left\{\begin{array}{cl}
\left|y_{s}^{(t)}(\rho)\right|  
& \text { when } \alpha_{\jmath_n}^{(t)} \leqslant \rho \leqslant \beta_{\jmath_n}^{(t)} 
\\ 
\left|b_{n_0}^{(t)}\right| 
& \text { when } \rho>\beta_{\jmath_n}^{(t)} 
\\ 
0 & \text { when } \rho<\alpha_{\jmath_n}^{(t)}
\end{array}\right.\end{array}
\label{eq24}
\end{equation}

Through the above analysis, it can be noted that for $s(t)$, its CWCF is a monotonically increasing linear function with a slope of 1 only when $t=0$, and at all other times, its CWCF is a periodic function. As for $\jmath(t)$, when $t=0$, its CWCF is a piecewise linear function with a linearly increasing segment having a slope of 1. When $t=t_{\jmath_n}$, its CWCF is an approximately piecewise linear function with an approximate slope less than 1. At all other times, its CWCF is approximately a periodic function.

\subsection{$y^{(t)}(\rho)$}
Upon careful examination of the aforementioned information, we can proceed to delve into the pertinent characteristics and effectiveness of $y^{(t)}(\rho)$. It is possible to express $y^{(t)}(\rho)$ as follows:
\begin{equation}
y^{(t)}(\rho) = y_{s}^{(t-\tau_s)}(\rho) + y_{\jmath}^{(t-\tau_\jmath)}(\rho)
\label{eq25}
\end{equation}

Furthermore, $\left|y^{(t)}\right|$ exhibits the following relation:
\begin{equation}
\begin{aligned}
\left||y_{s}^{(t-\tau_s)}|-|y_{\jmath}^{(t-\tau_{\jmath})}|\right|
\leqslant
|y^{(t)}|
\leqslant
\left||y_{s}^{(t-\tau_s)}|+|y_{\jmath}^{(t-\tau_\jmath)}|\right|
\end{aligned}
\label{eq26}
\end{equation}

Especially, when $t = \tau_s$, we have:
\begin{equation}
\begin{aligned}
\left|y^{(\tau_s)}\left(\rho\right)\right| = \left|y_{s}^{(0)}\left(\rho\right)+y_{\jmath}^{(\tau_s-\tau_\jmath)}\left(\rho\right)\right|
\end{aligned}
\label{eq27}
\end{equation} 

Based on (\ref{eq14}), (\ref{eq24}), (\ref{eq25}), and (\ref{eq27}), it can be deduced that $\left|y^{(\tau_s)}\left(\rho\right)\right|$ can be approximated as the combination of an autoterm, $\left|y_{s}^{(0)}\left(\rho\right)\right|$, and a crossterm, $\left|y_{\jmath}^{(\tau_s-\tau_\jmath)}\left(\rho\right)\right|$. In cases where $\tau_s-\tau_\jmath = t_{\jmath_n}$, the crossterm exhibits characteristics resembling those of a step function, thus significantly impeding the similarity between $\left|y^{(\tau_s)}\right|$ and $\left|y_s^{(0)}\right|$. However, when $\tau_s-\tau_\jmath \ne t_{\jmath_n}$, the crossterm can be approximated as a half-wave rectified function, with its magnitude diminishing as $|\tau_s-\tau_\jmath|$ increases. At this point, due to the periodic characteristics of the half-wave rectification function, we observe that $\left|y^{(\tau_s)}(\frac{T}{2})\right|\approx \left|y_s^{(0)}(\frac{T}{2})\right|$.

Similarly, when $t = t_{\jmath_n}+\tau_\jmath$, we have:
\begin{equation}
\begin{aligned}
\left|y^{(t_{\jmath_n}+\tau_\jmath)}\left(\rho\right)\right| = \left|y_{s}^{(t_{\jmath_n}+\tau_\jmath-\tau_s)}\left(\rho\right)+y_{\jmath}^{(t_{\jmath_n})}\left(\rho\right)\right|
\end{aligned}
\label{eq28}
\end{equation} 

From (\ref{eq13}), (\ref{eq20}), (\ref{eq25}), and (\ref{eq28}), it becomes apparent that the magnitude of $\left|y_{\jmath}^{(t_{\jmath_n}+\tau_\jmath)}\left(\rho\right)\right|$ can be approximated as an auto-term $\left|y_{\jmath}^{(t_{\jmath_n})}\left(\rho\right)\right|$, augmented by a cross-term $\left|y_{s}^{(t_{\jmath_n}+\tau_\jmath-\tau_s)}\left(\rho\right)\right|$. And there is $\left|y^{(t_{\jmath_n}+\tau_\jmath)}(\frac{T}{2})\right|\approx \left|y_\jmath^{(t_{\jmath_n})}(\frac{T}{2})\right|$.

For values of $t$ that do not satisfy $t = \tau_s$ or $t = t_{\jmath_n}+\tau_\jmath$, it can be deduced from (\ref{eq13}), (\ref{eq24}), and (\ref{eq25}) that the magnitude $\left|y^{(t)}(\rho)\right|$ corresponds to a complex envelope, devoid of the distinctive amplitude characteristics exhibited by $\left|y^{(\tau_s)}\left(\rho\right)\right|$ and $\left|y^{(t_{\jmath_n}+\tau_\jmath)}\left(\rho\right)\right|$. Furthermore, its maximum magnitude is significantly inferior to them.

After the above analysis, we describe the variation of the integrated energy of $v^{(t)}(\mu)$ in the waveform domain through $|y^{(t)}(\rho)|$. It is evident that only when $t=\tau_s$, $|y^{(\tau_s)}(\rho)|$ can be approximately viewed as a linear function. In comparison, when $t=\tau_{\jmath_n}$, $|y^{(\tau_{\jmath_n})}(\rho)|$ can be approximately seen as a stepwise linear function. At all other times, $|y^{(t)}(\rho)|$ can be regarded as a periodic function. Therefore, through statistical methods, we can determine whether each element on $|y^{(t)}(\rho)|$ is equivalent to the corresponding element of a linear function defined in the waveform domain. This helps identify effective and ineffective integration elements in the waveform domain at that moment. Consequently, we can establish the objective function as follows:
\begin{equation}
\begin{aligned}
O^{(t)}(\rho)
&=o^{(t)}\cdot\left(\rho+\frac{T}{2}\right)
\\
&=\frac{\left|y^{(t)}\left(\frac{T}{2}\right)\right|}{T}\cdot \left(\rho+\frac{T}{2}\right)
\end{aligned}
\label{eq29}
\end{equation}
where $o^{(t)} = \frac{\left|y^{(t)}\left(\frac{T}{2}\right)\right|}{T}$.

It is evident that when the cross-term becomes zero, and when $t=\tau_s$, the objective function $O^{(\tau_s)}(\rho)$ is equivalent to $\left|y^{(\tau_s)}(\rho)\right|$ itself. Conversely, when $t = t_{\jmath_n}+\tau_\jmath$, the approximate linear segments in $\left|y^{(t_{\jmath_n}+\tau_\jmath)}(\rho)\right|$ have a slope at least $\frac{T_J}{T_\jmath}$ times that of $o^{(t_{\jmath_n}+\tau_\jmath)}$, thereby exhibiting a distinct and discernible characteristic.

In the scenario where the cross-term is non-zero and there is additive Gaussian white noise, the problem of suppressing ISRJ can be formulated as a hypothesis testing problem, aiming to evaluate the equality between $\left|y^{(t)}(\rho)\right|$ and $O^{(t)}(\rho)$.

\section{Adaptive filtering term and compensation term}

In the subsequent sections, we will employ statistical methods to derive the adaptive weight $a^{(t)}(\mu)$ and the compensation term $\varsigma^{(t)}$ in the WD-AMF by leveraging the disparities between $\left|y^{(t)}(\rho)\right|$ and $O^{(t)}(\rho)$ based on statistical analysis.

\subsection{Noise model}
Assuming the existence of Gaussian white noise, which possesses additive characteristics represented as $\xi(t)\sim(0,\sigma^2)$ in the time domain, when substituting $\xi(t)$ into (\ref{eq6}) and (\ref{eq7}), the following results are obtained:
\begin{subequations}
\begin{equation}
\begin{aligned}
wgn^{(t)}(\mu) = \xi(t-\mu) h(\mu)
\end{aligned}
\end{equation}
\begin{equation}
bn^{(t)}(\rho) = \int_{-\infty}^{\rho} wgn^{(t)}(\mu) d \mu 
\end{equation}
\label{eq31}
\end{subequations}

Clearly, the term denoted as $wgn^{(t)}(\mu)$ maintains its characteristic as additive Gaussian white noise, adhering to the properties of $wgn^{(t)}(\mu)\sim(0,\sigma^2)$. Conversely, $bn^{(t)}(\rho)$ represents a standard Brownian noise, adhering to the properties of $bn^{(t)}(\mu)\sim\left[0,\left(\mu+\frac{T}{2}\right)\sigma^2\right]$, which is a Gauss-Markov random process.

\subsection{Filtering model}

Based on the analysis in the preceding sections, the anti-ISRJ problem can be mathematically formulated as a hypothesis testing problem to examine the equality between $\left|y^{(t)}(\rho)\right|$ and $O^{(t)}(\rho)$. However, the complex structure of $|y^{(t)}(\rho)|$ implies that different analytical solutions exist at different times $t$, making it unfeasible to use a single criterion for mathematical modeling.

To overcome this issue, an equivalent mathematical model is proposed as a probabilistic hypothesis testing problem of whether $|v^{(t)}(\rho)|=o^{(t)}$. 

If we denote the set of instances without jamming signal on $\mu$ as $\mathrm{U}^{(t)}_s$, when only a slice jamming is present on $\mu$ as $\mathrm{U}^{(t)}_\jmath$, and when both an echo signal and a slice jamming are present on $\mu$ as $\mathrm{U}^{(t)}_{s+\jmath}$, then the following relationships hold:
\begin{equation}
\begin{aligned}
&\frac{|v^{(\tau_s)}(\mu)|}{o^{(\tau_s)}}
\\
&=\left\{\begin{array}{cl}
1=\mathcal{A}^{(\tau_s)},&\quad\text { when } \mu \in \mathrm{U}^{(t)}_s
\\
\mathcal{M}^{(\tau_s)}(\mu),&\quad\text { when } \mu\in \mathrm{U}^{(t)}_{s+\jmath}
\end{array}\right.
\end{aligned}
\label{eq33}
\end{equation}
where $\mathcal{M}^{(\tau_s)}(\mu)\in \left[\left|1-\frac{A_\jmath}{A_s}\right|,1+\frac{A_\jmath}{A_s}\right]$, and
\begin{equation}
\begin{aligned}
&\frac{|v^{(t_{\jmath_n}+\tau_\jmath)}(\mu)|}{o^{(t_{\jmath_n}+\tau_\jmath)}} 
\\
&=
\left\{\begin{array}{cl}
\frac{T_J}{T_\jmath}\cdot\frac{1}{Q}\cdot \frac{A_s}{A_\jmath}=\mathcal{C}^{(t_{\jmath_n}+\tau_\jmath)}, &\quad\text { when } \mu \in \mathrm{U}^{(t)}_s
\\
\frac{T_{J}}{T_\jmath}=\mathcal{A}^{(t_{\jmath_n}+\tau_\jmath)}, &\quad\text { when } \mu\in \mathrm{U}^{(t)}_\jmath
\\
\mathcal{M}^{(t_{\jmath_n}+\tau_\jmath)}(\mu), &\quad\text { when } \mu\in \mathrm{U}^{(t)}_{s+\jmath}
\\
0,&\quad\text { else }
\end{array}\right.
\end{aligned}
\label{eq34}
\end{equation}
where $\mathcal{M}^{(t)}(\mu)\in \left[\left|\mathcal{A}^{(t)}-\mathcal{C}^{(t)}\right|,\mathcal{A}^{(t)}+\mathcal{C}^{(t)}\right]$.

(\ref{eq33}) and (\ref{eq34}) imply that if $|v^{(t)}(\mu)| = o^{(t)}$, then $t = \tau_s$ and $\mu \in \mathrm{U}^{(t)}_s$. Additionally, for self-defensive forwarding interference, we observe $\mathcal{A}^{(t_{\jmath_n}+\tau_\jmath)}>2$, meaning that when $\mu\in \mathrm{U}^{(t)}_\jmath$, we have $|v^{(t_{\jmath_n}+\tau_\jmath)}(\mu)|>2o^{(t_{\jmath_n}+\tau_\jmath)}$. As previously stated, our focus centers around $\left|v^{(t)}(\mu)\right|$ for $\mu \in \mathrm{U}^{(t)}_\jmath, \mathrm{U}^{(t)}_{s+\jmath}$. Specifically, our objective is to preserve $v^{(\tau_s)}(\mu)$ to the greatest extent while minimizing $v^{(t_{\jmath_n}+\tau_\jmath)}(\mu)$. 

It is evident that when $A_{\jmath}\gg A_s$, we have $\left|\mathcal{A}^{(t)}-\mathcal{C}^{(t)}\right|>2$, leading to $\left|v^{(t)}(\mu)\right|>2o^{(t)}$ for $\mu \in \mathrm{U}^{(t)}_\jmath, \mathrm{U}^{(t)}_{s+\jmath}$. However, in scenarios where $A_{\jmath}$ is relatively small, it is possible to encounter $\left|\mathcal{A}^{(t)}-\mathcal{C}^{(t)}\right|<2<\mathcal{A}^{(t)}+\mathcal{C}^{(t)}$, and thus the condition $\left|v^{(t)}(\mu)\right|>2o^{(t)}$ for $\mu \in \mathrm{U}^{(t)}_\jmath, \mathrm{U}^{(t)}_{s+\jmath}$ is not universally valid. Considering that $\mathcal{M}^{(t)}(\mu)$ is a continuous function within the range $\left[\left|\mathcal{A}^{(t)}-\mathcal{C}^{(t)}\right|, \mathcal{A}^{(t)}+\mathcal{C}^{(t)}\right]$, and the intervals $\mathrm{U}^{(t)}_\jmath$ and $\mathrm{U}^{(t)}_{s+\jmath}$ are relatively short, we can extend $\mu$ to obtain $v^{(t)}(\mu\pm\gamma\cdot d\mu)>2o^{(t)}$, for $\mu \in \mathrm{U}^{(t)}_\jmath, \mathrm{U}^{(t)}_{s+\jmath}$, where $\gamma$ deones the scaling factor.

Therefore, the anti-ISRJ problem can be reformulated as a probabilistic hypothesis testing problem, determining whether $|v^{(t)}(\mu\pm\gamma\cdot d\mu)|>2o^{(t)}(\mu)$. Consequently, the anti-ISRJ problem can be transformed into an unbiased estimation problem for $v^{(t)}(\mu)$ and $y^{(t)}(\rho)$. We denote their estimates as $\hat v^{(t)}(\mu)$ and $\hat y^{(t)}(\rho)$, respectively.

\subsection{State estimation model}

In scenarios where $t=\tau_s$ and $t=\tau_\jmath$, and the impact of crossterm can be deemed insignificant, it is feasible to approximate $y^{(t)}(\mu)$ as locally exhibiting a linear relationship of first order. This characteristic enables us to model $\hat y^{(t)}(\mu)$ by employing a linear function model complemented by two impulse function models.

In the subsequent steps, our objective is to establish models for $\hat y^{(t)}(\mu)$ and $\hat v^{(t)}(\mu)$ using the Interactive Multiple Model Kalman Filter algorithm (IMM-KF)\cite{23}. Within the framework of the IMM-KF algorithm, the interdependent state of $\hat y^{(t)}(\mu)$ can be precisely defined as a weighted combination of three distinct model states, given by the expression:
\begin{equation}
\hat M^{(t)}(\mu|\mu) = u_1\hat M^{(t)}_1(\mu|\mu) + u_2\hat M^{(t)}_2(\mu|\mu) + u_3\hat M^{(t)}_3(\mu|\mu)
\label{eq35}
\end{equation}
Here, the probabilities $u_1$, $u_2$, and $u_3$ are determined for each model based on the residuals and residual covariance obtained through the utilization of the Kalman filter. $\hat M^{(t)}_1(\mu|\mu)$, $\hat M^{(t)}_2(\mu|\mu)$, and $\hat M^{(t)}_3(\mu|\mu)$ represent the estimated states of their respective models, and their one-step prediction state equations are expressed as follows:
\begin{subequations}
\begin{equation}
\begin{aligned}
&\hat M^{(t)}_1(\mu+d\mu|\mu)
= F_1\hat M^{(t)}(\mu|\mu)
\\
&=\left[\begin{array}{ccccc}
1 & d\mu & 0 & 0 & 0\\  0 & 1 & 0 & 0 & 0\\ 0 & 0 & 0 & 0 & 0\\ 0 & 0 & 0 & 1 &0 \\ 0 & 0 & 0& 0 & 1
\end{array}\right]
\left[\begin{array}{l}
\hat y^{(t)}(\mu) \\ \hat v^{(t)}(\mu) \\ \hat \delta_{-}^{(t)}(\mu) \\ \hat \delta_{+}^{(t)}(\mu) \\ wgn^{(t)}(\mu) \end{array}\right]
\end{aligned}
\end{equation}
\begin{equation}
\begin{aligned}
&\hat M^{(t)}_2(\mu+d\mu|\mu)
=F_2 \hat M^{(t)}(\mu|\mu)
\\
&=\left[\begin{array}{ccccc}
1 & d\mu & d\mu & 0 & 0\\  0 & 1 & d\mu & 0 & 0\\ 0 & -1 & 0 & 0 & 0\\ 0 & 0 & 0 & 1 & 0\\ 0& 0 & 0 &0 & 1
\end{array}\right]
\left[\begin{array}{l}
\hat y^{(t)}(\mu) \\ \hat v^{(t)}(\mu) \\ \hat \delta_{-}^{(t)}(\mu) \\ \hat \delta_{+}^{(t)}(\mu) \\ wgn^{(t)}(\mu)\end{array}\right]
\end{aligned}
\end{equation}
\begin{equation}
\begin{aligned}
&\hat M^{(t)}_3(\mu+d\mu|\mu)
= F_3 \hat M^{(t)}(\mu|\mu)
\\
&=\left[\begin{array}{ccccc}
1 & d\mu & 0 & d\mu & 0\\  0 & 1 & 0 & d\mu & 0\\ 0 & 0 & 0 & 0 & 0\\ 0 & 0 & 0 & 1 & 0\\ 0 & 0 & 0 &0 &1
\end{array}\right]
\left[\begin{array}{l}
\hat y^{(t)}(\mu) \\ \hat v^{(t)}(\mu) \\ \hat \delta_{-}^{(t)}(\mu) \\ \hat \delta_{+}^{(t)}(\mu)\\wgn^{(t)}(\mu) \end{array}\right]
\end{aligned}
\end{equation}
\label{36}
\end{subequations}
Here, $F_i$, with $i=1,2,3$, denotes the matrices governing state transitions. The entities $\hat \delta_{-}^{(t)}(\mu)$ and $\hat \delta_{+}^{(t)}(\mu)$ correspond to distinct impulse functions exerting influence over both the direction and magnitude of $\hat v^{(t)}(\mu)$. It is noteworthy that the state matrix has been expanded in this context due to the measurement value $y^{(t)}(\mu)$ satisfying the Brownian noise model.

For the model $\hat M_1^{(t)}$, $\hat v^{(t)}(\mu+d\mu|\mu)$ is a constant. This model describes the linear integration of $\hat y^{(t)}(\mu)$ with a fixed $\hat v^{(t)}(\mu)$.

For the model $\hat M_2^{(t)}$, the value of $\hat v^{(t)}(\mu+d\mu|\mu)$ undergoes a linear variation induced by $\hat \delta_{-}^{(t)}(\mu)$. It is assumed that the negative impulse function $\hat \delta_{-}^{(t)}(\mu)d\mu$ is incorporated into $\hat y_x^{(t)}(\mu+d\mu|\mu)$. As a result, an impulse function $\hat \delta_{-}^{(t)}(\mu+d\mu|\mu) = -\hat v^{(t)}(\mu)$ emerges. This model effectively captures the sudden transition process in $\hat v^{(t)}$ and $\hat y^{(t)}(\mu)$ when the signal dissipates.

Regarding the model $\hat M_3^{(t)}$, the value of $\hat v^{(t)}(\mu+d\mu|\mu)$ experiences a linear variation caused by $\hat \delta_{+}^{(t)}(\mu)$. It is postulated that the positive impulse function $\hat \delta_{+}^{(t)}(\mu)d\mu$ is incorporated into $\hat y^{(t)}(\mu+d\mu|\mu)$. Consequently, the impulse function remains constant, denoted as $\hat \delta_{+}^{(t)}(\mu+d\mu|\mu) = \hat \delta_{+}^{(t)}(\mu)$. This model effectively captures the abrupt transition process in $\hat v^{(t)}$ and $\hat y^{(t)}(\mu)$ when the signal emerges.

Since the constructed model is a Markov model, we are unable to derive variable $\hat \delta_{+}^{(t)}(\mu)$ from the state equation. However, based on prior analysis, we may infer that $\hat \delta_{+}^{(t)}(\mu)$ is equal to $K\cdot E^{(t)},K>2$. 

Indeed, $\hat M_1^{(t)}$ constitutes a substantial proportion of $\hat M^{(t)}$, considering that only a negligible fraction of time corresponds to high weights of $\hat M_2^{(t)}$ and $\hat M_3^{(t)}$. Hence, the probability transition matrix can be represented as follows:
\begin{equation}
\begin{aligned}
P^{(t)} =
\left[\begin{array}{ccc}
1-2p_0 & p_0 & p_0\\  1 & 0 & 0 \\ 1 &0 &0
\end{array}\right]
\end{aligned}
\label{38}
\end{equation}
where $p_0$ represents the probability of a sudden change in the $|v^{(t)}|$, and the zero elements in the diagonal of the matrix are not strictly zero, but typically represent a very small value to ensure matrix invertibility.

When $t$ does not equal $t_s+\tau_s$ or $t_\jmath+\tau_\jmath$, the complex variable $\hat v^{(t)}(\mu)$ ceases to remain constant. Although the IMM weighted state output may offer an approximation of $\hat v^{(t)}(\mu)$, the accuracy of estimating its absolute value gradually diminishes as $\hat \delta_{+}^{(t)}(\mu)$ decreases. As a result, the state estimation of $\hat y^{(t)}(\mu)$ and $\hat v^{(t)}(\mu)$, $\mu\in \mathrm{U}_\jmath,\mathrm{U}_{s+\jmath}$ becomes biased. 

\subsection{$a^{(t)}(\mu)$ and $\varsigma^{(t)}$}
Given that $\hat y^{(t)}(\mu)$ is a biased estimation, we continue to use $o^{(t)}$ as the decision criterion. Thus, we define the adaptive threshold as:
\begin{equation}
E^{(t)}(\mu) = 2o^{(t)}
\end{equation}
and the adaptive weight function $a^{(t)}(\mu)$ can be represented as:
\begin{subequations}
\begin{equation}
\begin{array}{l}
\mathrm{Y}\left[y^{(t)}(\mu)\right] = a^{(t)}(\mu)
\\
=\left\{\begin{array}{l}
0,\quad\text { when } \mu \in \mathrm{U}^{(t)}_v
\\
1,\quad\text { when } \mu \in \mathrm{U}^{(t)}_e
\end{array}\right.\end{array}
\label{eq40}
\end{equation}
\begin{equation}
\mathrm{U}^{(t)}_v = \left\{\mu\pm \gamma \cdot d\mu\bigg|\left|\hat v^{(t)}(\mu)\right|>E^{(t)}(\mu)\right\}
\end{equation}
\begin{equation}
\mathrm{U}^{(t)}_e = \mathrm{Cu}\mathrm{U}^{(t)}_{v}
\end{equation}
\end{subequations}
where $\mathrm{U}^{(t)}_e$ represents the set of effective integration elements, and $\mathrm{U}^{(t)}_v$ represents the set of ineffective integration elements.

Let the length of $\mathrm{U}^{(t)}_{v}$ be $L_v$, and the length of $\mathrm{U}^{(t)}_{e}$ be $L_e$. Suppose $\Psi^{(t)}$ is a random subset of $\mathrm{U}^{(t)}_{e}$ with a length of $L_v$. Then $\varsigma^{(t)}$ can be expressed as:
\begin{equation}
\begin{aligned}
\varsigma^{(t)} = \mathrm{V}\left[y^{(t)}(\mu)\right]= \int_{\Psi^{(t)}} \hat v^{(t)}(\mu) \mathrm{d}\mu
\end{aligned}
\label{eq44}
\end{equation}
Simultaneously, without loss of generality, since $\left|\hat{v}^{(t)}(\mu)\right| \gg E^{(t)}(\mu)$ when $t\ne\tau_s$ and $t\ne\tau_s+t_{\jmath_n}$, leading to $\mathrm{U}^{(t)}_e$ being an empty set, it is necessary to compensate for the noise energy in the output. Therefore, the final expression for WD-AMF output $z_o(t)$ can be represented as:
\begin{equation}
\begin{aligned} 
&z_{o}(t) 
\\
& =\int_{-\infty}^{\infty} a^{(t)}(\mu) v^{(t)}(\mu) d \mu+\varsigma^{(t)} + \int_{\Phi^{(t)}} wgn_{c}^{(t)}(\mu)
\end{aligned}
\label{eq47}
\end{equation}
where $\Psi^{(t)}$ represents a random subset of length $L_v$ in the waveform domain. $wgn_{c}^{(t)}$ denotes Gaussian white noise with the same distribution as $wgn^{(t)}$ but is statistically independent.
\begin{figure*}[htbp]
\centering
\subfloat[]{
\includegraphics[width=5.5 cm]{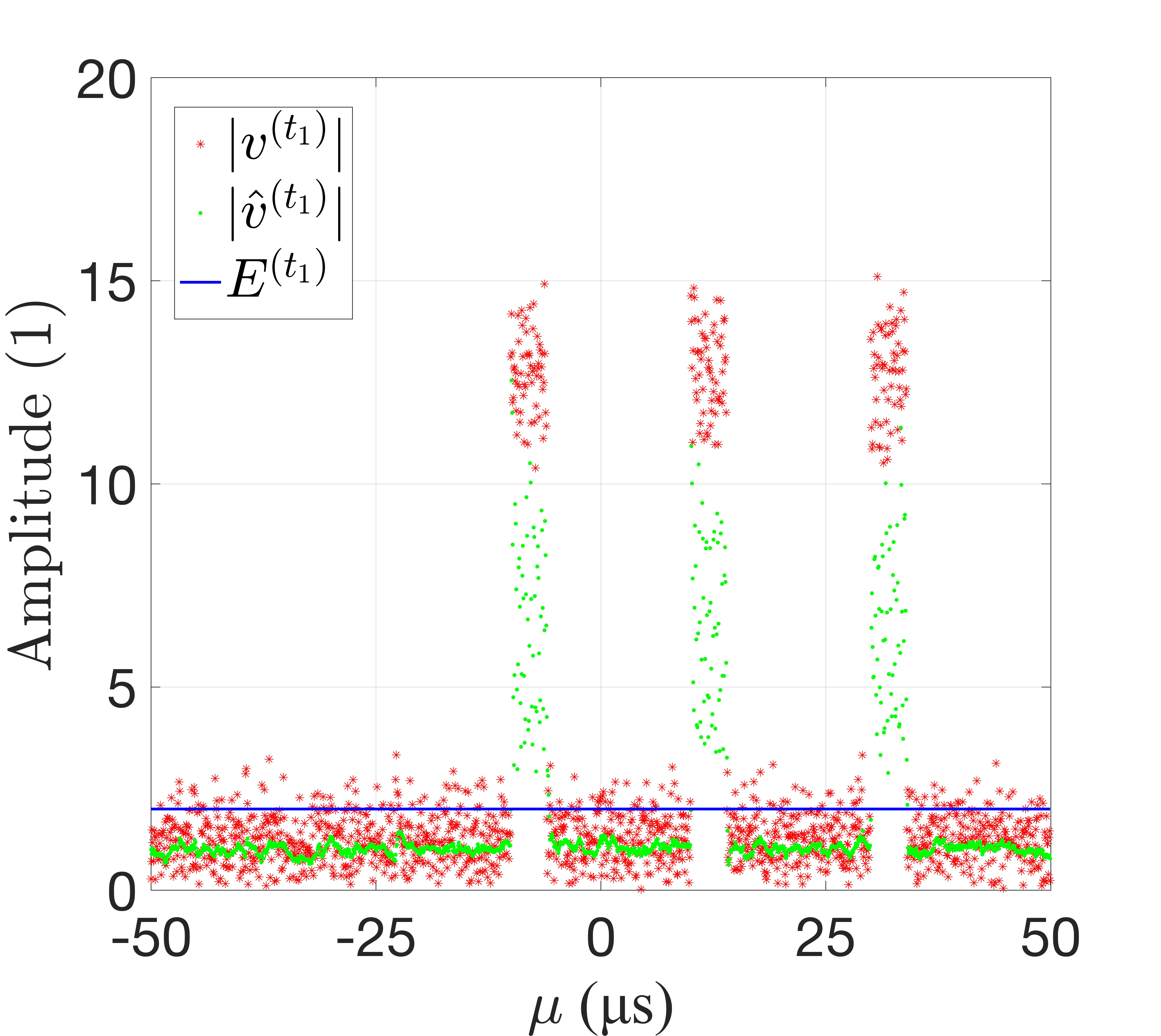}
\label{fig1(a)}}%
\subfloat[]{
\includegraphics[width=5.5 cm]{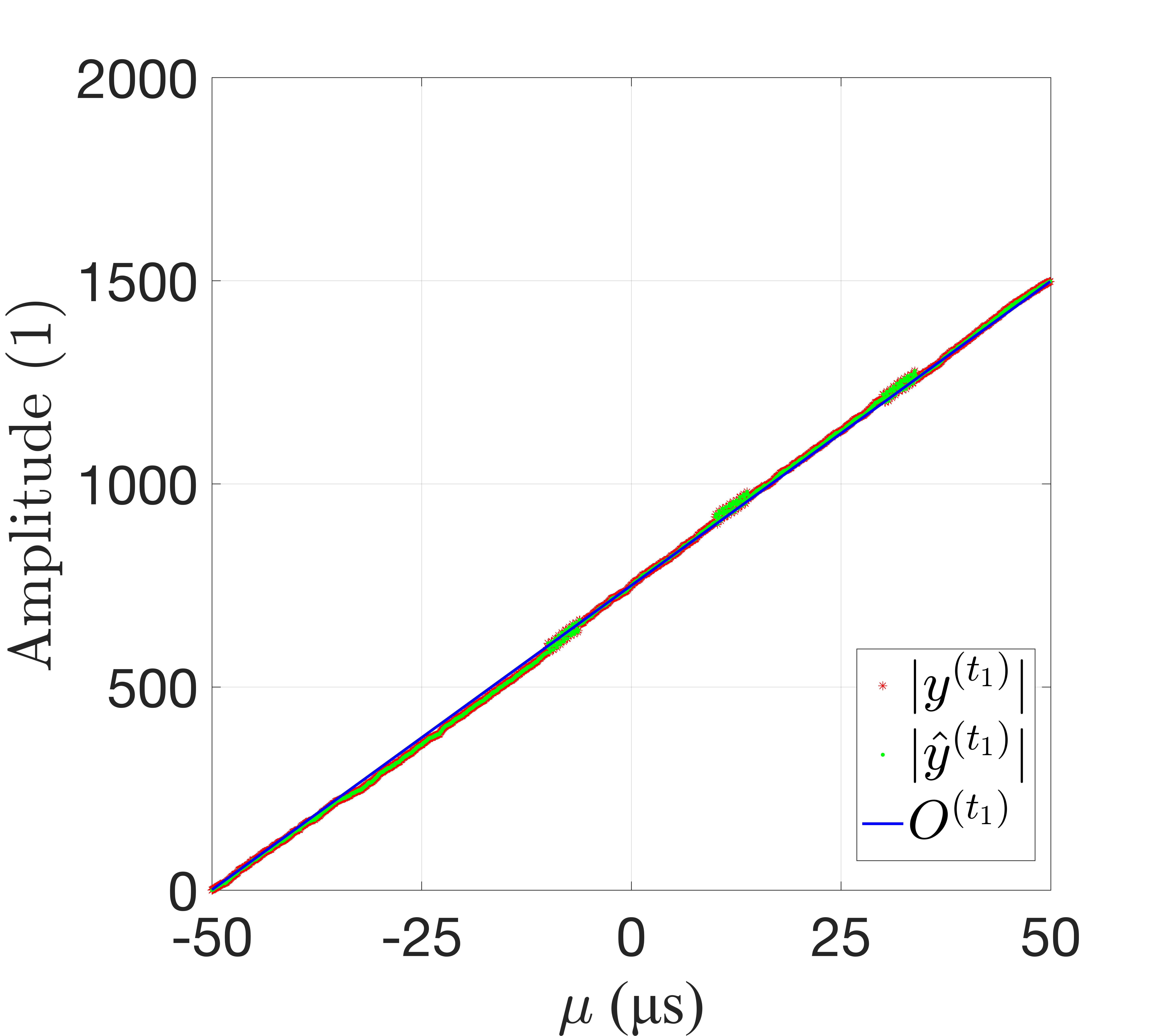}
\label{fig1(b)}}%
\subfloat[]{
\includegraphics[width=5.5 cm]{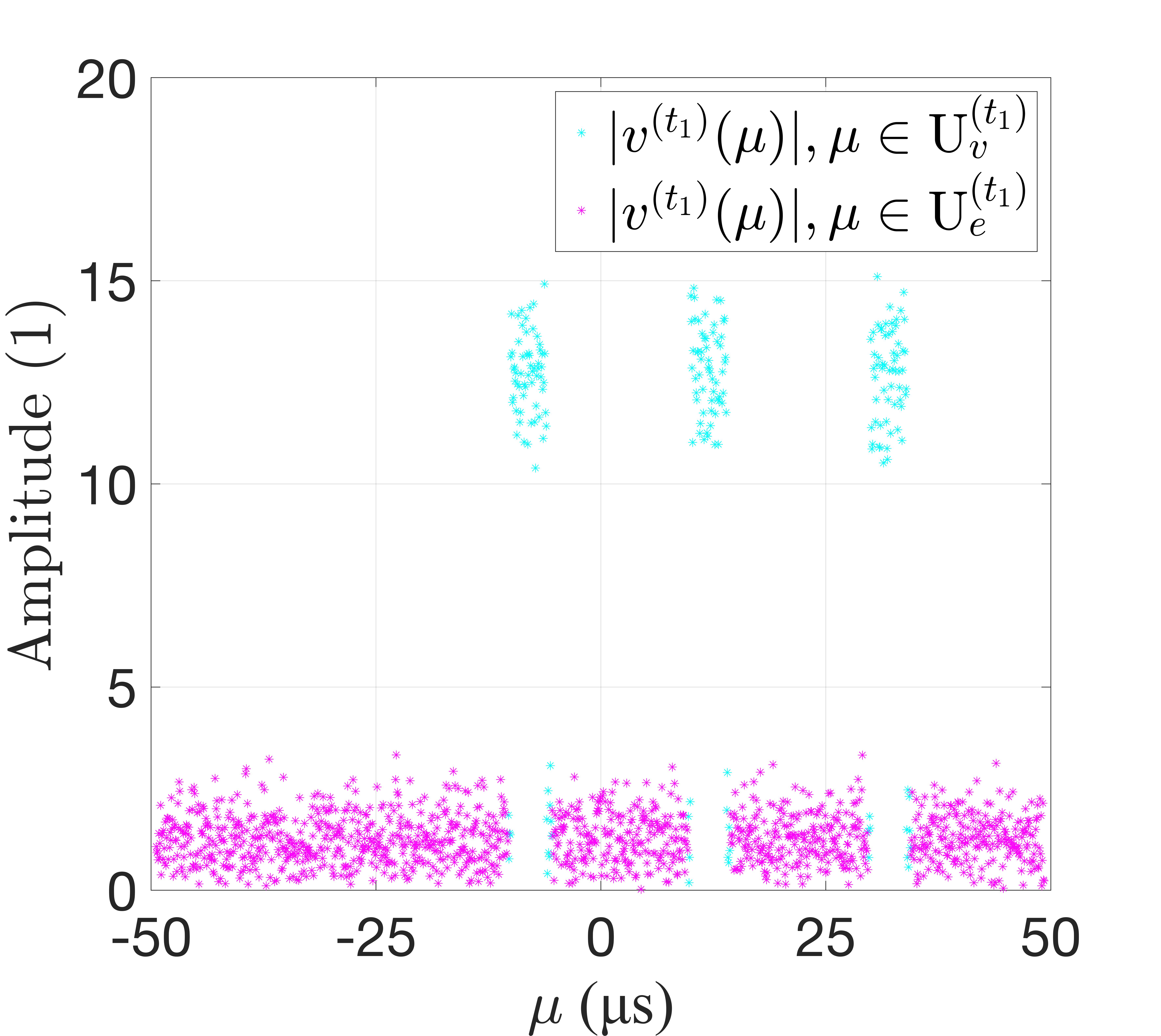}
\label{fig1(c)}}%
\\
\subfloat[]{
\includegraphics[width=5.5 cm]{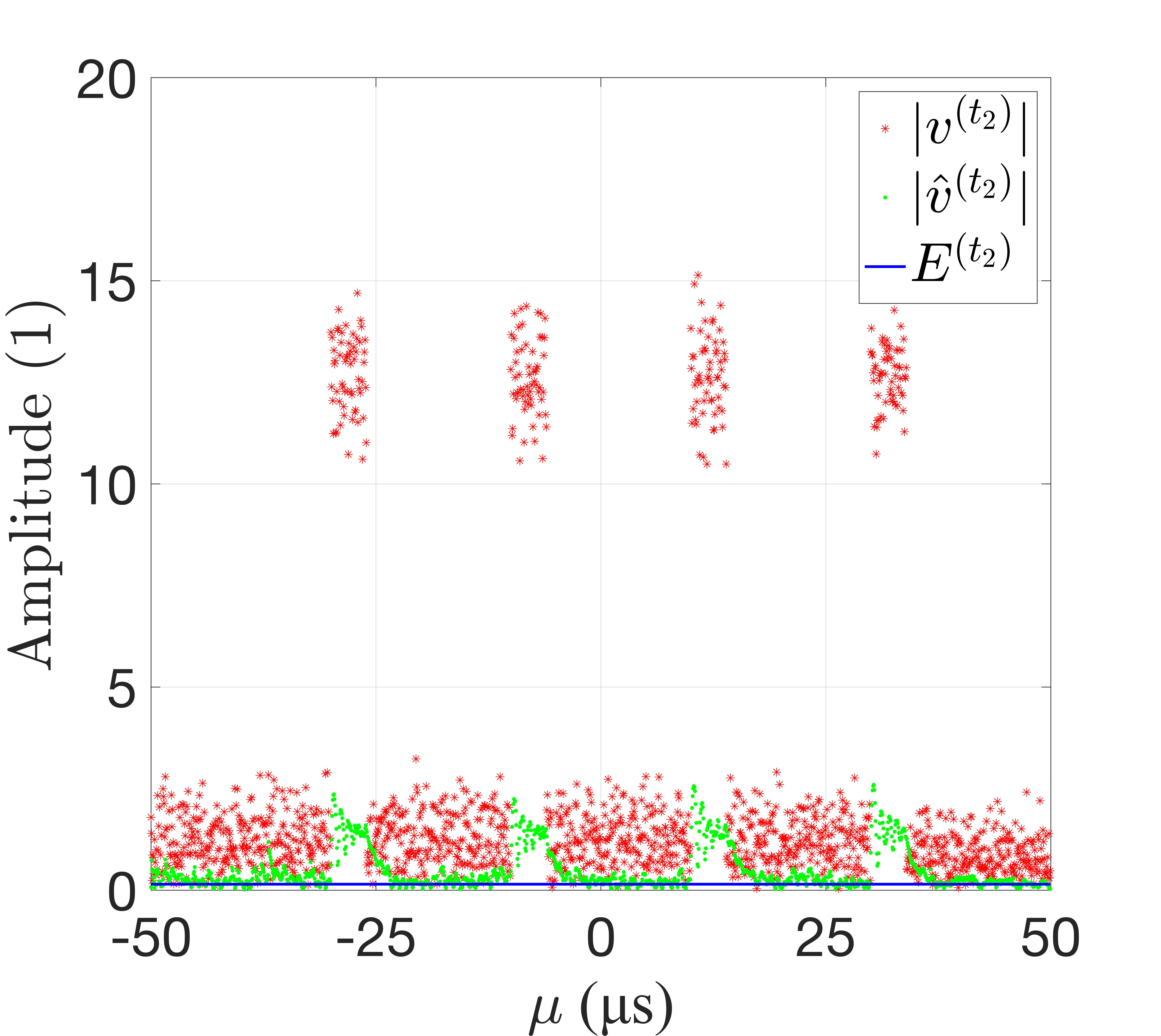}
\label{fig1(d)}}%
\subfloat[]{
\includegraphics[width=5.5 cm]{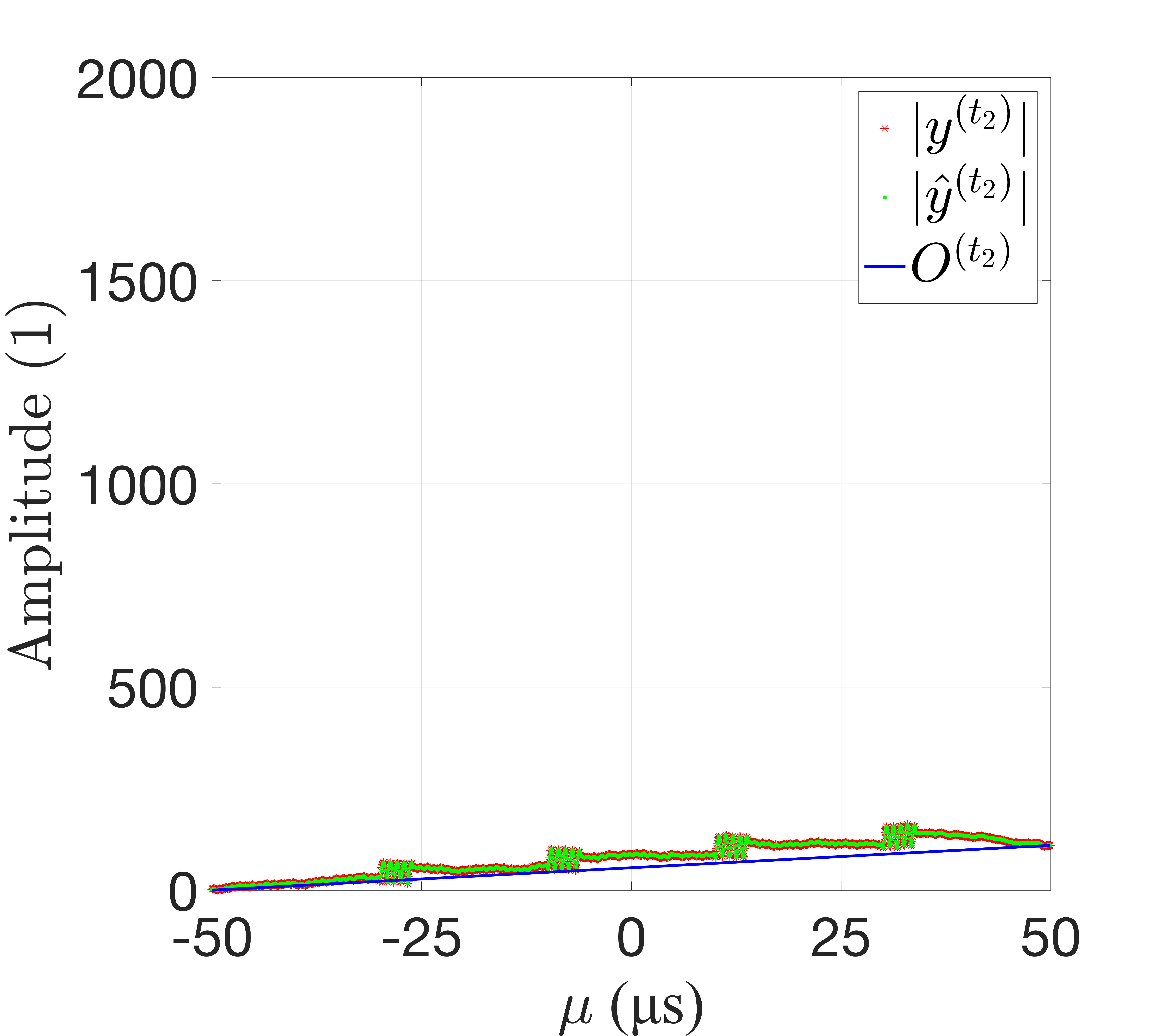}
\label{fig1(e)}}%
\subfloat[]{
\includegraphics[width=5.5 cm]{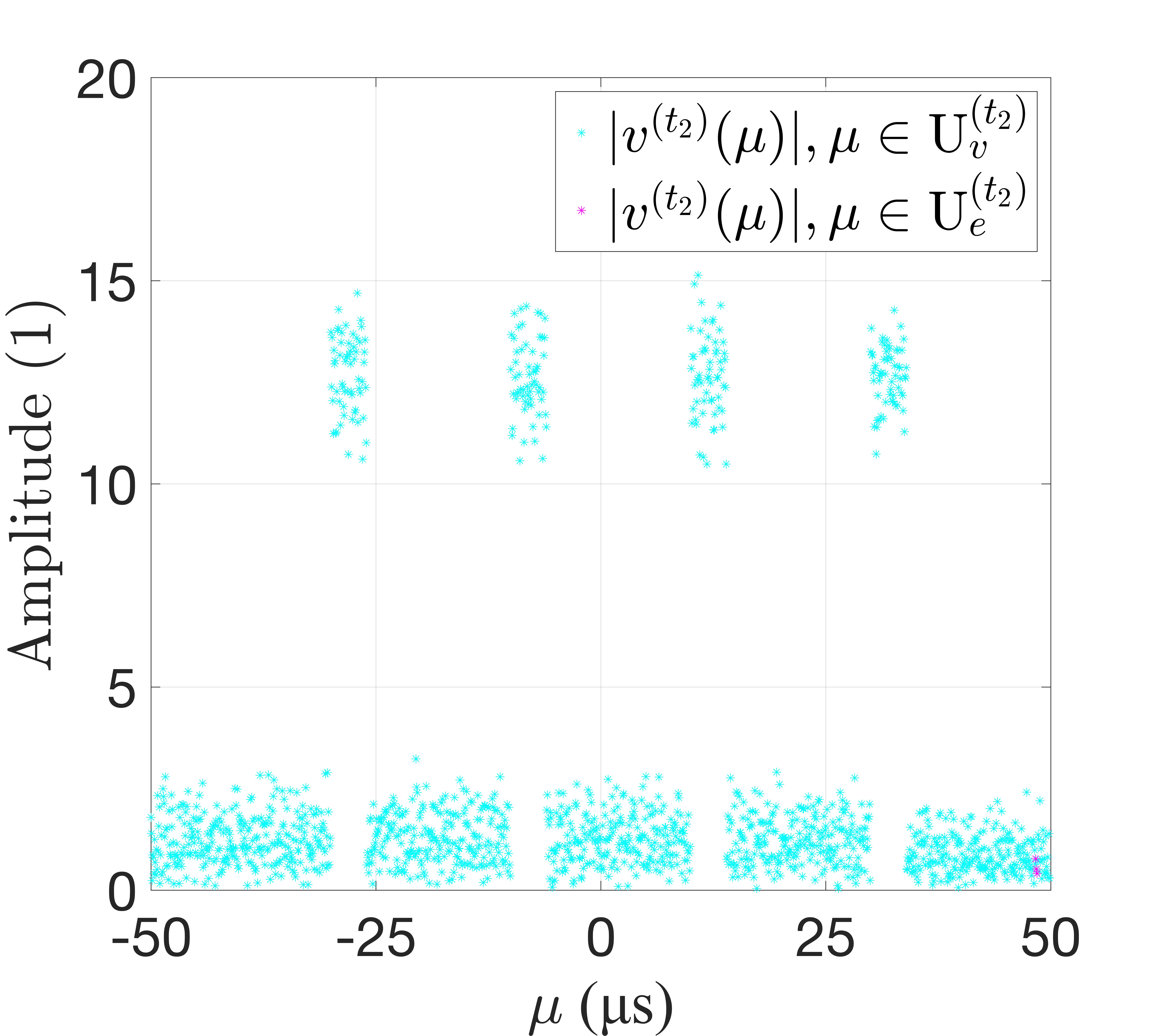}
\label{fig1(f)}}%
\\
\subfloat[]{
\includegraphics[width=5.5 cm]{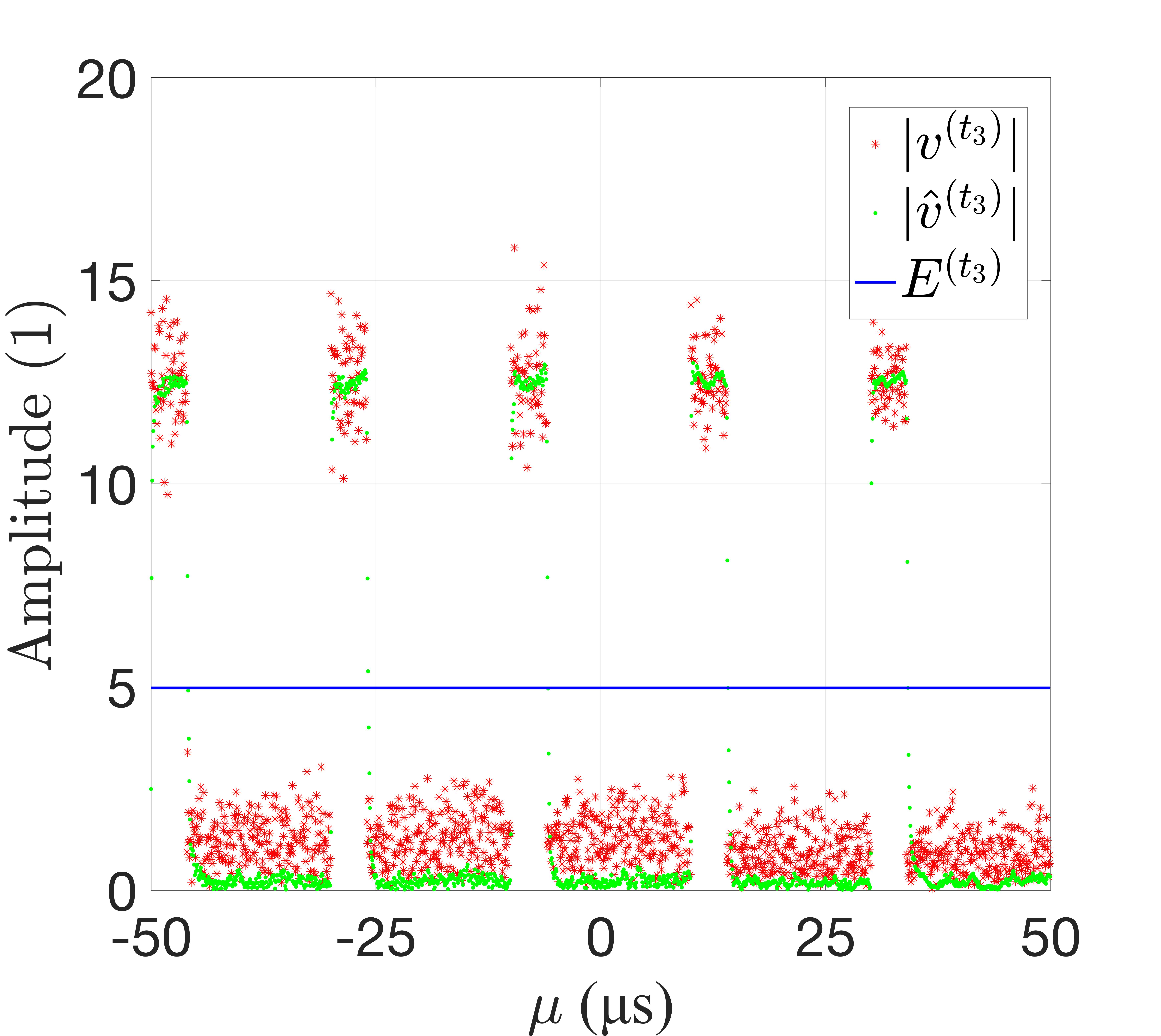}
\label{fig1(g)}}%
\subfloat[]{
\includegraphics[width=5.5 cm]{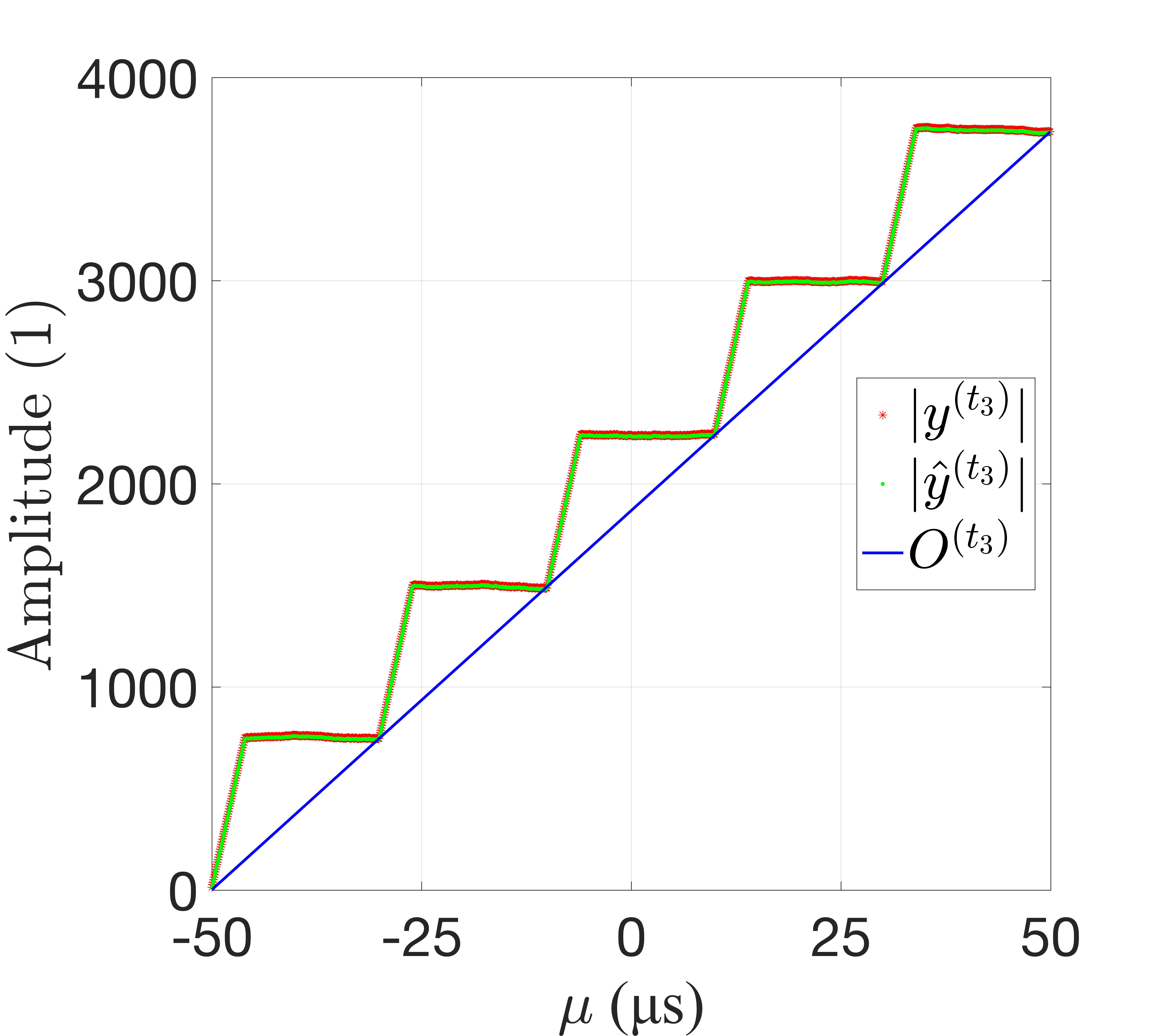}
\label{fig1(h)}}%
\subfloat[]{
\includegraphics[width=5.5 cm]{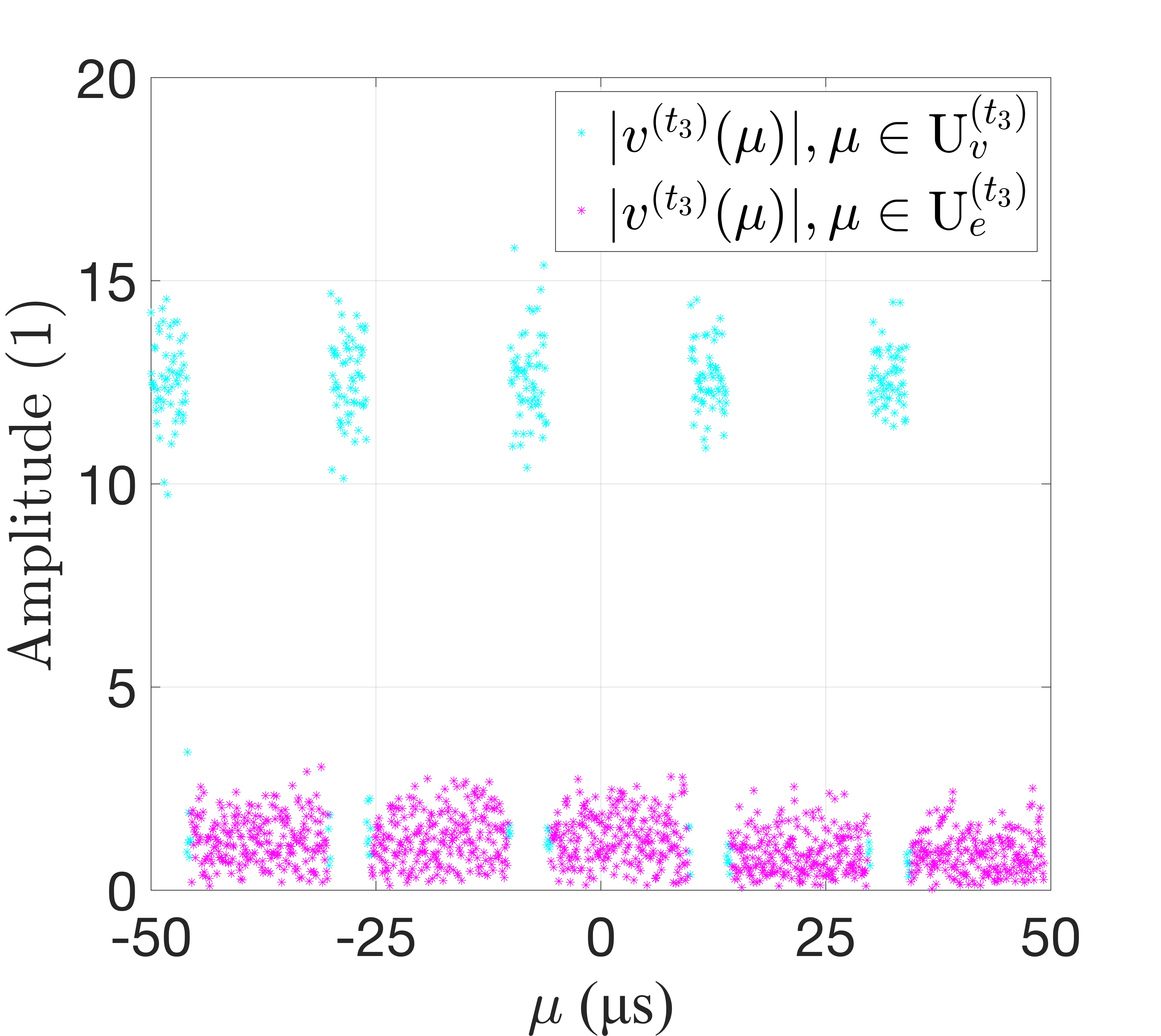}
\label{fig1(i)}}%
\centering
\caption{The simulated intermediate results (outputs of IMM-KF) of WD-AMF for $x(t)$ at time $t_1$, $t_2$, and $t_3$, under SNR = $0$ dB, SJR = $-15$ dB. (a) State estimation $|\hat v^{(t)}(\mu)|$ at $t_1$; (b) State estimation $|\hat y^{(t)}(\mu)|$ at $t_1$; (c) Labeling results at $t_1$. (d) State estimation $|\hat v^{(t)}(\mu)|$ at $t_2$; (e) State estimation $|\hat y^{(t)}(\mu)|$ at $t_2$; (f) Labeling results at $t_2$. (g) State estimation $|\hat v^{(t)}(\mu)|$ at $t_3$; (h) State estimation $|\hat y^{(t)}(\mu)|$ at $t_3$; (i) Labeling results at $t_3$. 
\label{fig1}}
\end{figure*}

\section{Numerical examples}

In this section, numerical illustrations are employed to validate the efficacy of the proposed approach.

\subsection{Construction of WD-AMF}
\label{S6A}
Let us assume that the transmitter employs a baseband LFM waveform with a pulsewidth of $100$ $\upmu$s and a bandwidth of $6$ MHz. The receiver operates at a sampling frequency of $15$ MHz. The SNR is set at $0$ dB. Furthermore, the interrupted-sampling frequency is specified as $50$ KHz, with the slice width to interrupted-sampling repeater period ratio of $\varepsilon=\frac{T_\jmath}{T_J}=0.2$. It is important to note that the ISRJ experiences a time delay, denoted as $\tau_\jmath - \tau_s$, of $40$ $\upmu$s relative to the echo signal. The SJR is established at $-15$ dB.

Considering the moment when the echo signal emerges as time zero, we designate $t_1=0$, $t_2=20$ $\upmu$s, and $t_3=40$ $\upmu$s. Fig. \ref{fig1} illustrates the simulated intermediate results (outputs of IMM-KF) of the WD-AMF method at these three time points.

Fig. \ref{fig1}\subref{fig1(a)}-\ref{fig1}\subref{fig1(c)} depict the simulated intermediate results (outputs of IMM-KF) of WD-AMF of $x(t)$ at time $t_1$. Fig. (\ref{fig1(a)}) and Fig. \ref{fig1}\subref{fig1(b)} illustrate the estimated values of the waveform-domain states $|\hat v^{(t)}(\mu)|$ and $|\hat y^{(t)}(\mu)|$, respectively, where the red markers denote the measured values and the green markers denote the estimated values. The black solid lines in Fig. \ref{fig1}\subref{fig1(a)} and Fig. \ref{fig1}\subref{fig1(b)} represent the adaptive threshold $E^{(t)}$ and the objective function $O^{(t)}$, respectively. The simulation results show that the IMM-KF algorithm provides a good estimation of the part $\mu\in\mathrm{U}^{(t)}_e$, but a large deviation occurs in the estimation of the part $\mu\in\mathrm{U}^{(t)}_v$. This is because when $A_\jmath>\hat \delta_{+}^{(t)}$, the IMM cannot describe this type of nonlinearity well. However, this does not affect the subsequent processing results. As analyzed earlier, even if some interference elements leak into our adaptive decision interval, their continuous integration value in the waveform domain is extremely small, so the impact on the final filtering result can be neglected. Fig. \ref{fig1}\subref{fig1(c)} shows the labeling results of the interference component and non-interference component in $|v^{(t)}(\mu)|$ obtained through the adaptive threshold $E^{(t)}$, indicating that both are well distinguished.

Fig. \ref{fig1}\subref{fig1(d)}-\ref{fig1}\subref{fig1(f)} display the simulated intermediate results (outputs of IMM-KF) of WD-AMF applied to $x(t)$ at time $t_2$. The obtained simulation outcomes confirm that the IMM-KF algorithm continues to provide a reliable estimation of the linear component of $|v^{(t)}(\mu)|$ and $|y^{(t)}(\mu)|$. Due to the minute magnitude of $E^{(t)}$, only an insignificantly small portion of the non-interference component enters the adaptive decision interval, as depicted in Fig. \ref{fig1}\subref{fig1(f)}.

Fig. \ref{fig1}\subref{fig1(g)}-\ref{fig1}\subref{fig1(i)} showcase the simulated intermediate results (outputs of IMM-KF) of WD-AMF applied to $x(t)$ at time $t_3$. The obtained simulation results demonstrate that the IMM-KF algorithm yields commendable estimations over the entire waveform domain. This favorable outcome arises due to the capability of the IMM to effectively capture this type of nonlinearity when $A_s\ll\hat \delta_{+}^{(t)}$ within the algorithm. Notably, Fig. \ref{fig1}\subref{fig1(i)} distinctly exhibits the differentiation between the interference component and non-interference component in $|v^{(t)}(\mu)|$, thereby enabling the exclusive integration of $|z_o(t_1)|$ over the non-interference component.
\begin{figure}[tbp]
\includegraphics[width=7 cm]{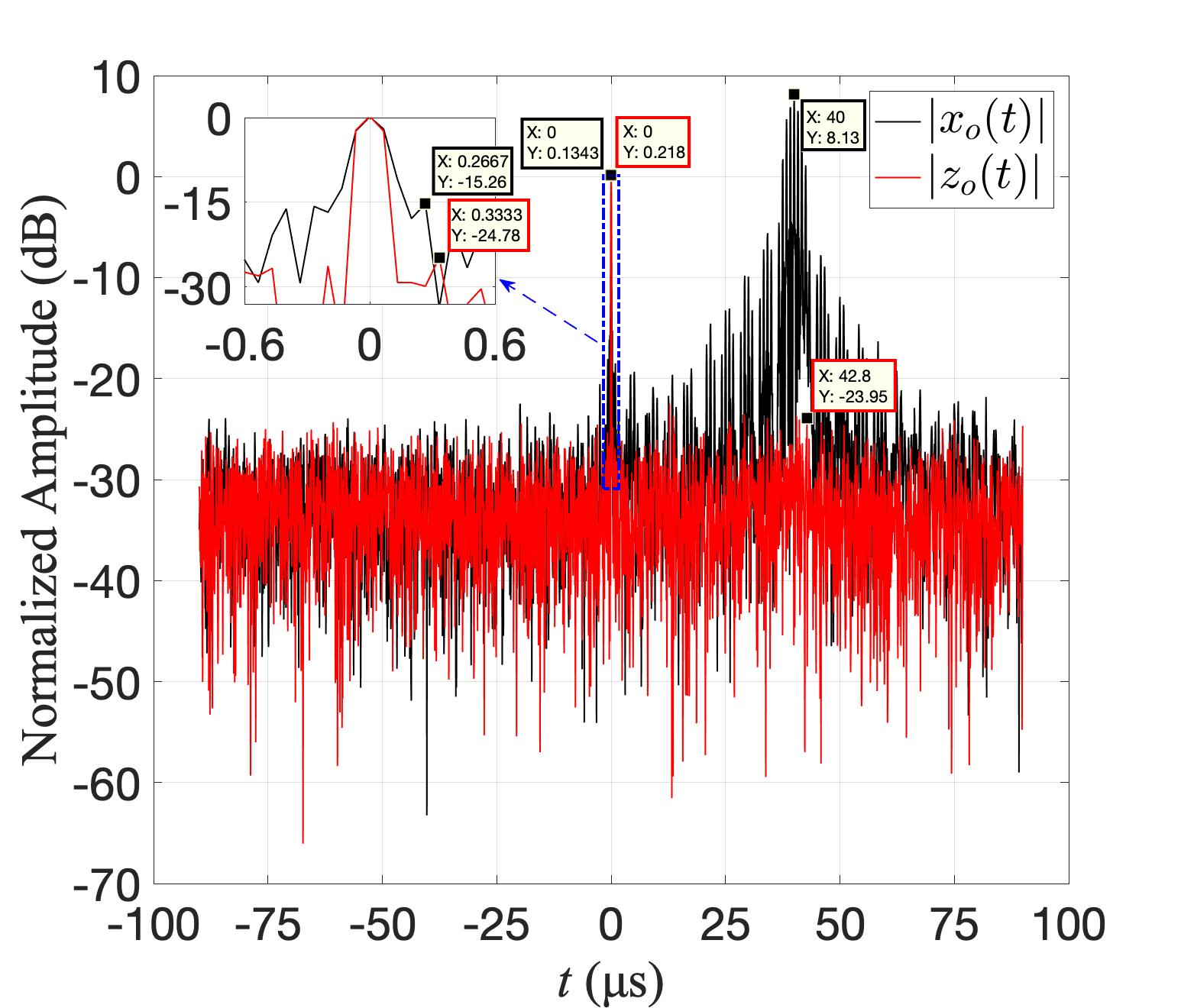}
\centering
\caption{The normalized amplitude output of the filters under SNR = $0$ dB and SJR = $-15$ dB.
\label{fig2}}
\end{figure}

Fig. \ref{fig2} presents the normalized amplitude output results obtained from WD-AMF. The black solid line corresponds to the output of the matched filter, denoted as $|x_o(t)|$, while the red line represents the output of WD-AMF, denoted as $|z_o(t)|$. Analysis of Fig. \ref{fig2} reveals that $|z_o(t)|$ effectively suppresses interference signals without compromising the amplification of the echo signal. Additionally, in comparison to the output results of matched filter, the output results of WD-AMF demonstrate a reduced peak level for the first sidelobe. Moreover, the SJR achieved after applying WD-AMF reaches a value of $24$ dB, accompanied by a significant $32$ dB suppression of interference signal gain by the matched filter.

\subsection{Evaluation of ISRJ Resistance}

This subsection aims to analyze the system's performance in the presence of a mobile point target and multiple sources of jamming. The simulation parameters of the jamming scene can be found in Tab \ref{tab1}. It is assumed that the two jamming sources share the same jamming characteristics. To facilitate a comparative analysis, the anti-ISRJ algorithms described in literature \cite{9} and literature \cite{20} have been selected. The LFM waveform parameters employed by the three algorithms are as follows: the bandwidth ($B$) is set to $6$ MHz, the pulse width ($T$) is set to $100$ $\upmu$s, the interrupted-sampling frequency ($f_\mathrm{s}$) is set to $100$ kHz, and $\varepsilon $ is set to $0.25$. Furthermore, for the algorithm introduced in literature \cite{9}, the SNR loss is assumed to be $1$ dB. It is important to note that both approaches presented in \cite{9} and \cite{20} necessitate prior knowledge of the interference signal's parameters. Hence, it is presumed that the parameters of the interfering signals in \cite{9} and \cite{20} are already known.
\begin{table}[tbp]
\fontsize{8}{10}\selectfont 
\caption{Simulation parameters of the jamming scene} 
\centering
\setlength{\tabcolsep}{3pt}
\begin{tabular}{p{95pt}<{\centering}p{100pt}<{\centering}}
\toprule
Parameters&Value\\
\hline
Radar carrier frequency      &$f_0$ = 2 GHz\\
Pulse repetition frequency        &PRF = 1 KHz\\
\hline   
Target radial distance        &$d_0$ = 60 km\\
Target radial velocity        &$\nu_0$ = 300 m/s\\
\hline   
Jamming radial distance        &$d_1$ = 54 km, $d_2$ = 120 km\\
Jamming radial velocity        &$\nu_1$ = -300 m/s, $\nu_2$ = 300 m/s\\
\hline
Signal to noise ratio        &SNR = 0 dB\\
Signal to jamming ratio        &SJR = -20 dB\\
\bottomrule
\end{tabular}
\label{tab1}
\end{table}
\begin{figure*}[htbp]
\centering
\subfloat[]{
\includegraphics[width=7 cm]{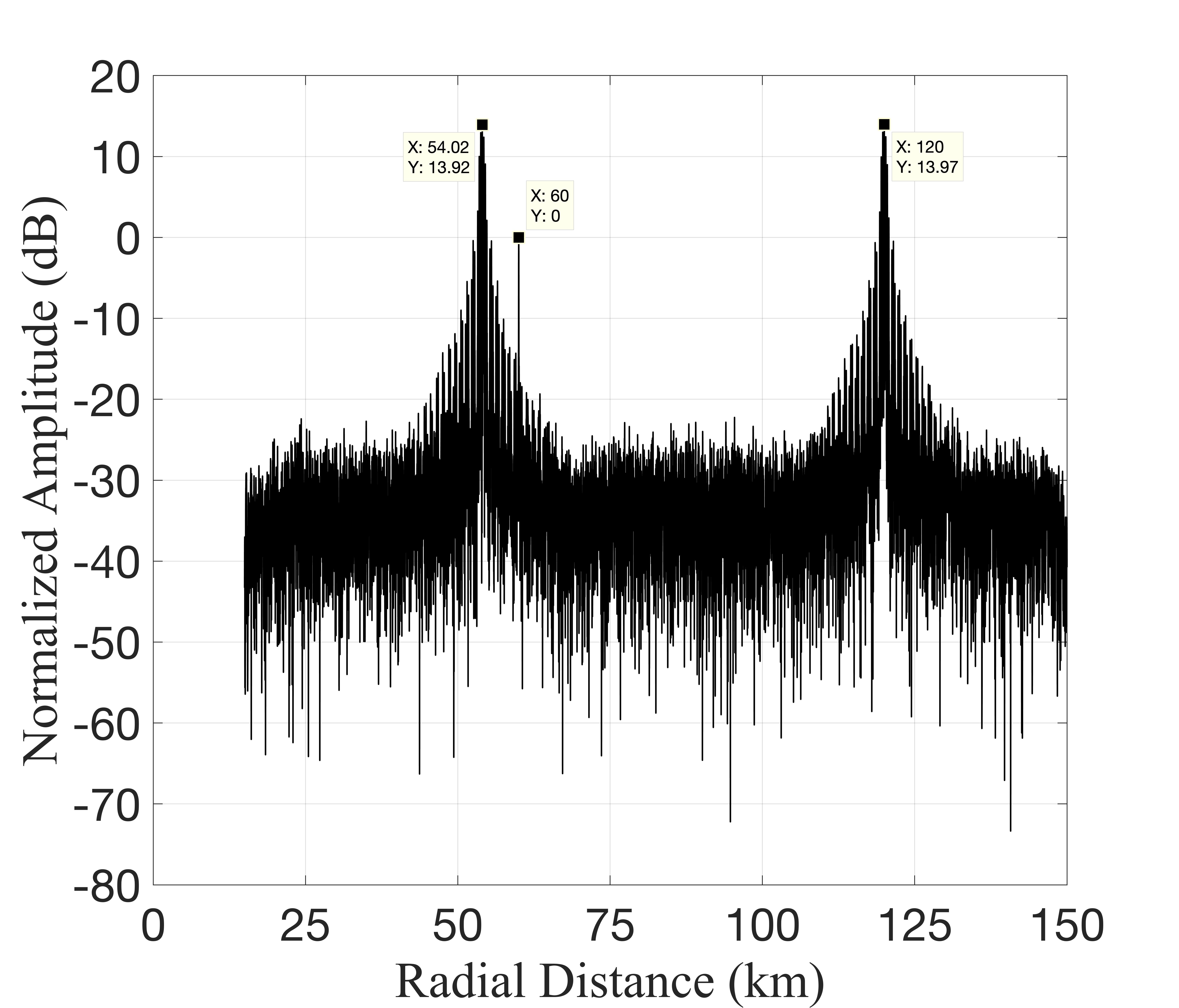}
\label{fig3(a)}}%
\subfloat[]{
\includegraphics[width=7 cm]{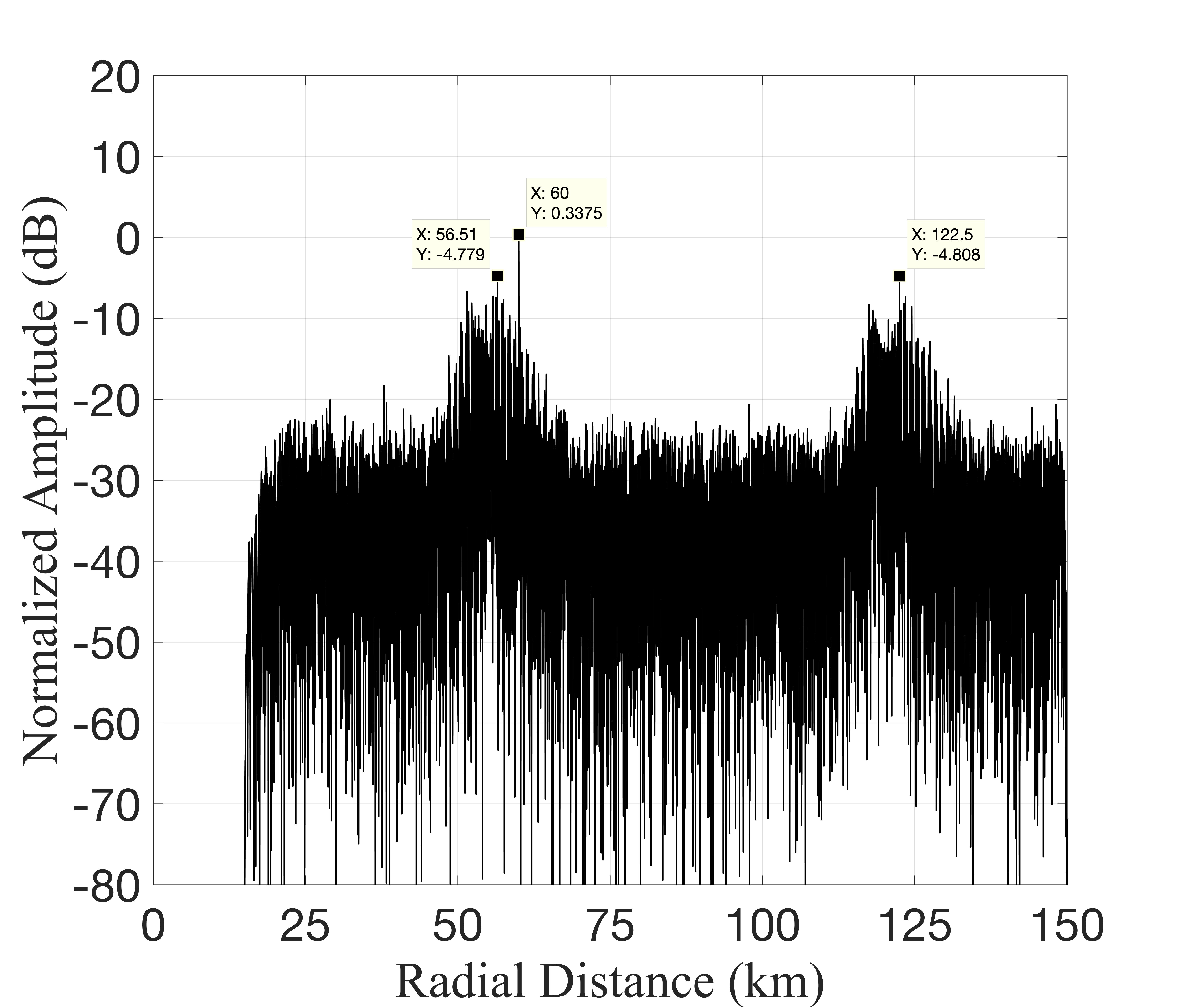}
\label{fig3(b)}}%
\\
\subfloat[]{
\includegraphics[width=7 cm]{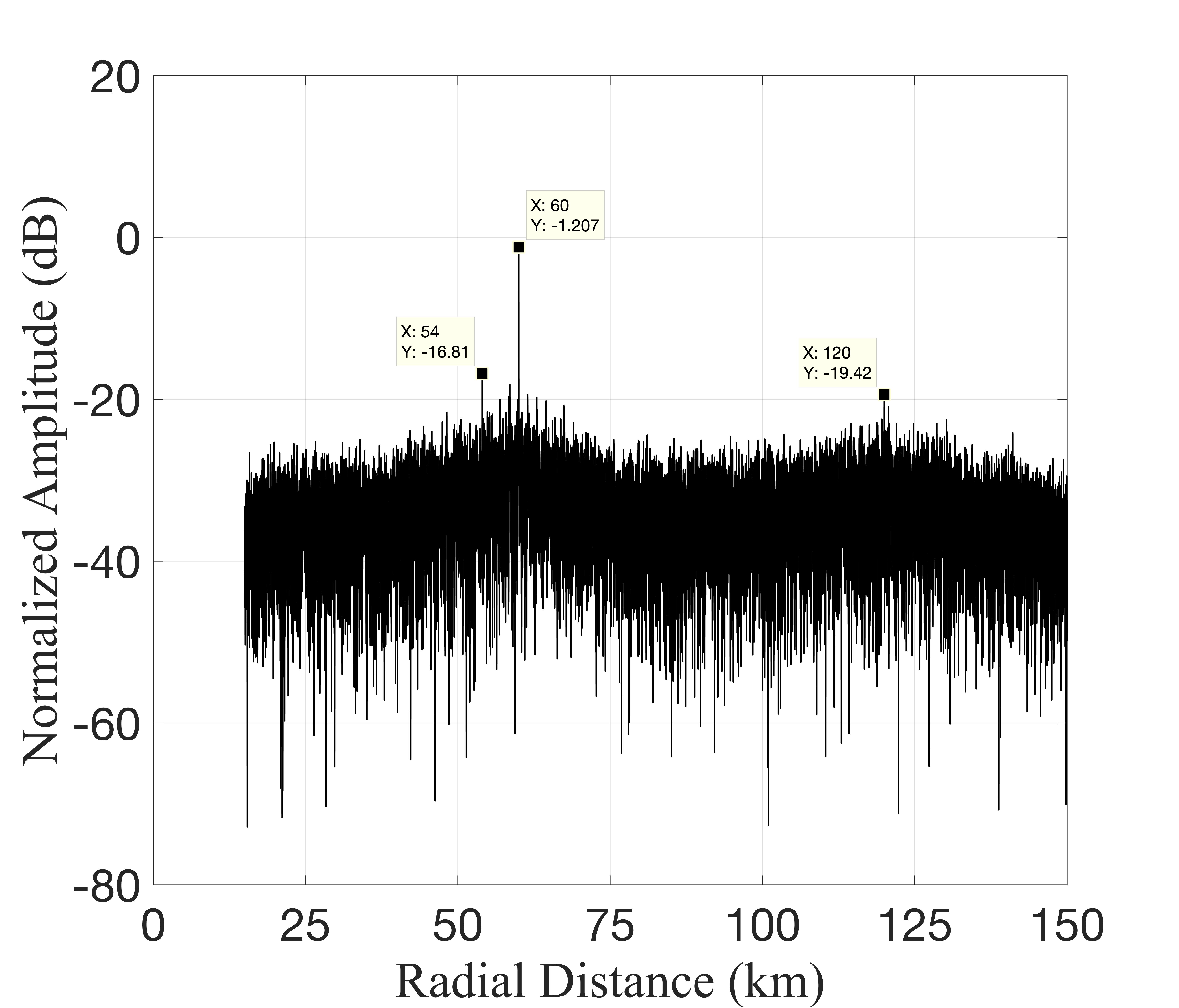}
\label{fig3(c)}}%
\subfloat[]{
\includegraphics[width=7 cm]{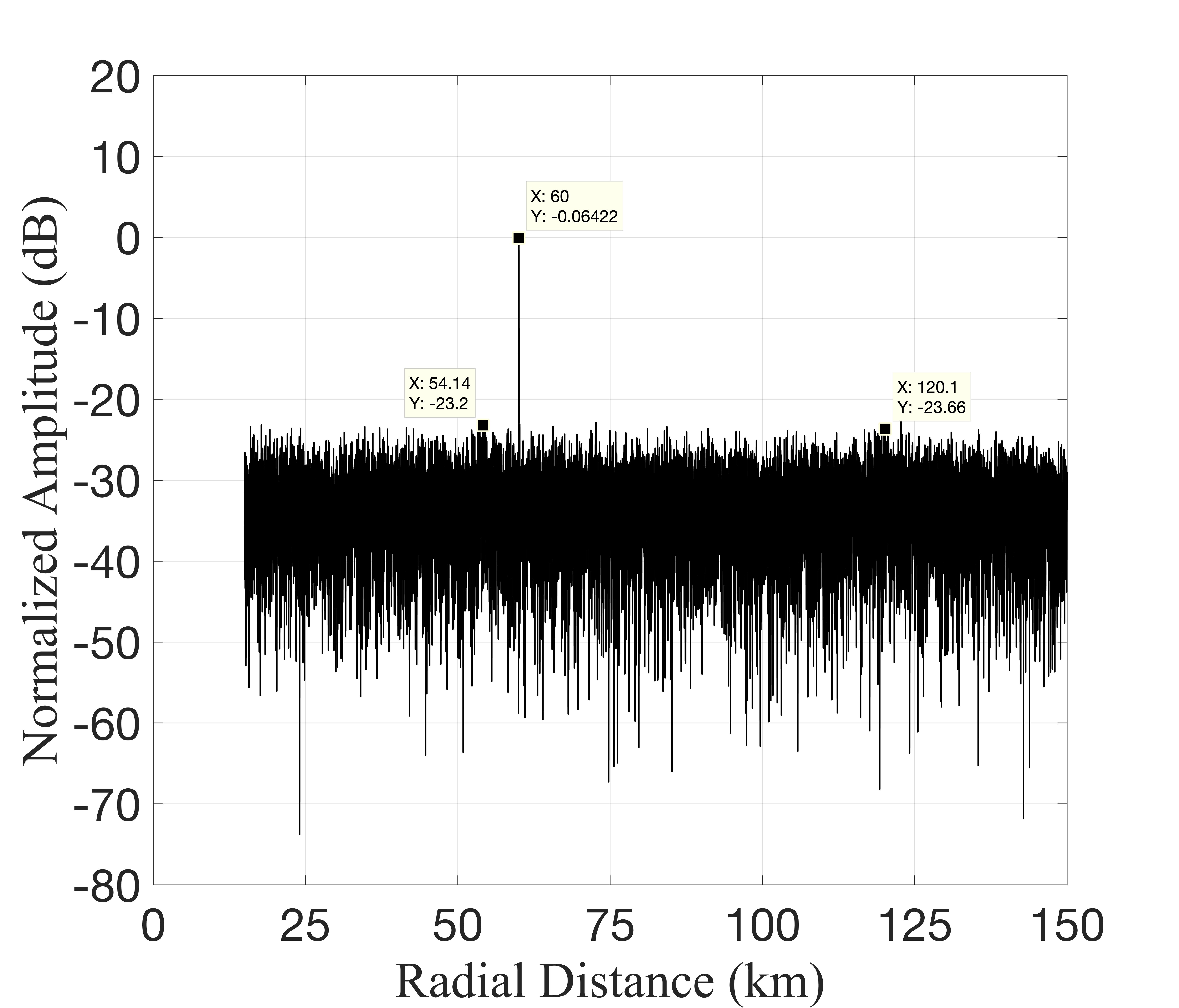}
\label{fig3(d)}}%
\centering
\caption{The output results of different methods in the interference scenario. (a) MF algorithm; (b) The time-frequency filtering method described in \cite{20}; (c) The Waveform-filter design method presented in \cite{9}; (d) WD-AMF method proposed in this paper.
\label{fig3}}
\end{figure*}

Fig. \ref{fig3} illustrates the output results of different algorithms, with Fig. \ref{fig3}\subref{fig3(a)} specifically displaying the output results of the MF algorithm. In the simulation scenario described in this paper, the approach introduced in literature \cite{20} yields a substantial number of spurious targets, greatly impairing the detection of weak targets. In contrast, the method presented in literature \cite{9} effectively mitigates false targets, exhibiting an approximate difference of $16$ dB between the peak of the interference output and the peak of the target output. Notably, our proposed method attains the lowest sidelobe level, showcasing an approximate difference of $23$ dB between the peak of the interference output and the peak of the target output.

\begin{figure*}[htbp]
\centering
\subfloat[]{
\includegraphics[width=7 cm]{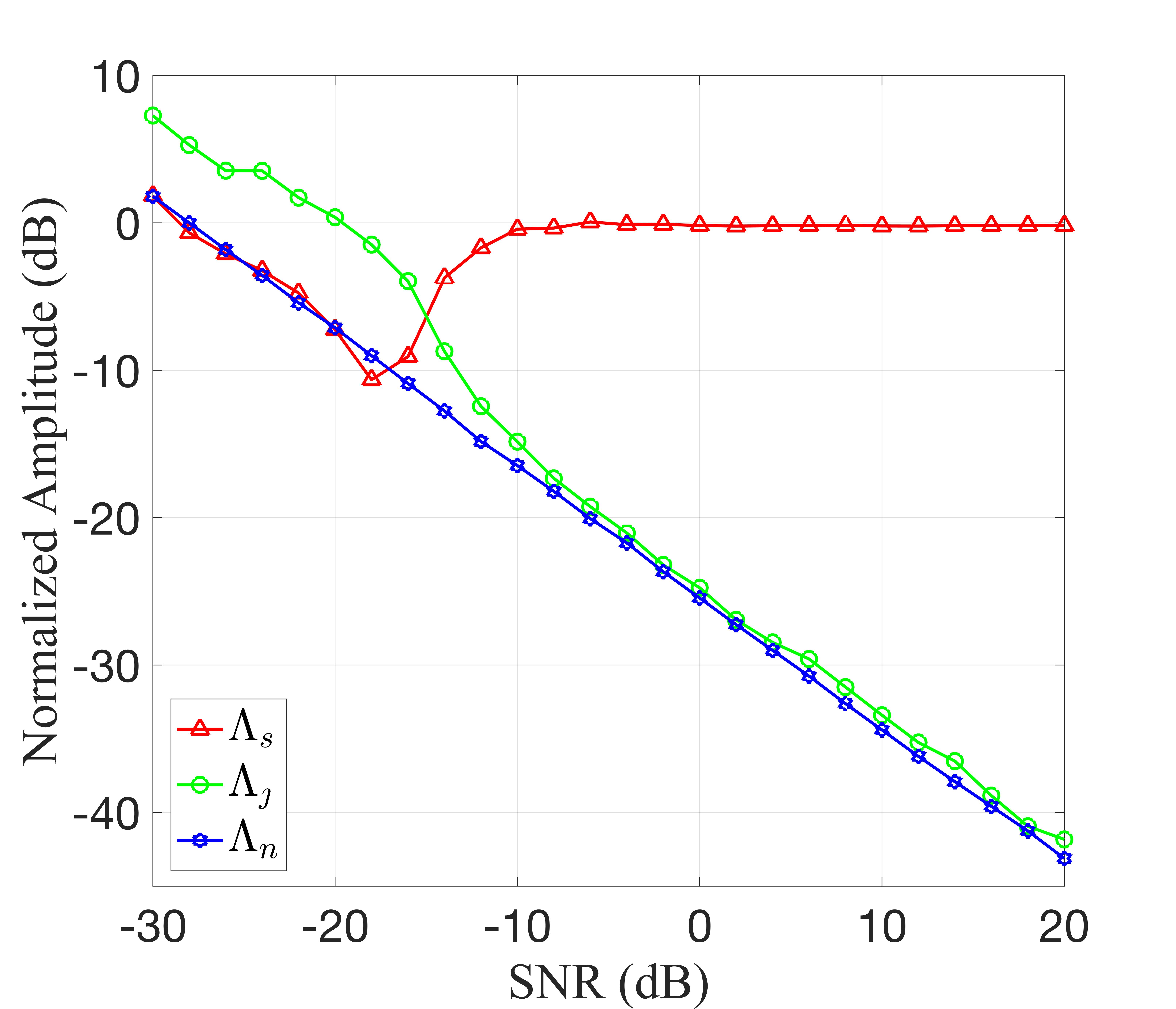}
\label{fig4(a)}}%
\subfloat[]{
\includegraphics[width=7 cm]{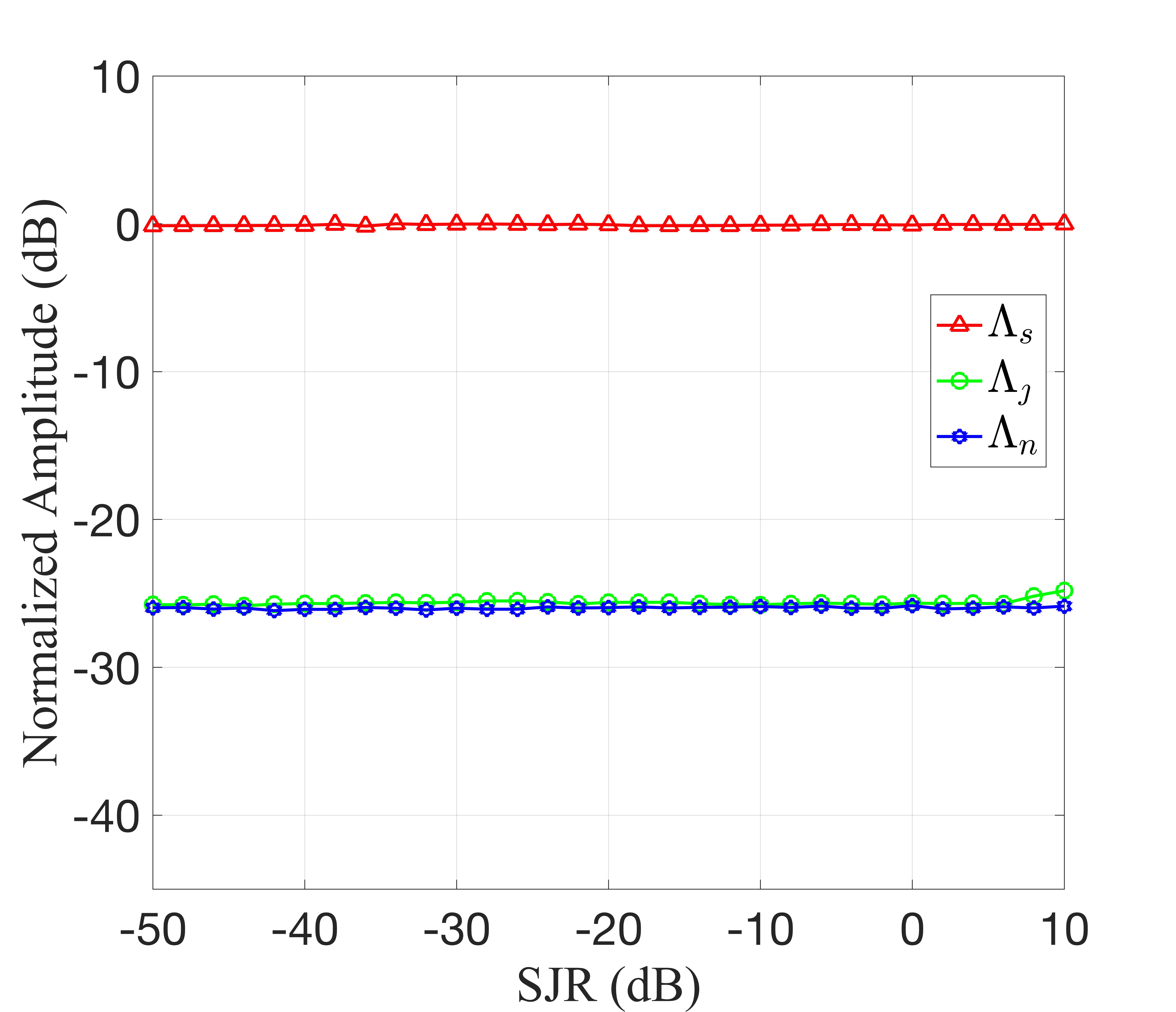}
\label{fig4(b)}}%
\centering
\caption{The graphs depict the variations in the levels of the target and interference peaks produced by the proposed method at different SNRs and SJRs. (a) The curves represent the changes in the levels of the target and interference peaks as a function of SNR; (b) The curves illustrate the fluctuations in the levels of the target and interference peaks as a function of SJR.
\label{fig4}}
\end{figure*}

We have conducted additional verification of the output performance of our proposed method under various SNRs and SJRs within the scenario presented in Tab \ref{tab1}. To mitigate the influence of noise randomness, we performed $200$ Monte Carlo simulations for each SNR and SJR parameter. Let $\Lambda_s$, $\Lambda_\jmath$, and $\Lambda_n$ respectively denote the average peak levels of the target, interference, and noise. Fig. \ref{fig4} illustrates the average target peak value across multiple simulations. It is important to note that, in Fig. \ref{fig4}\subref{fig4(a)}, the SJR has been fixed at $-20$ dB, while in Fig. \ref{fig4}\subref{fig4(b)}, the SNR has been fixed at $0$ dB.

From Fig. \ref{fig4}\subref{fig4(a)}, it can be inferred that when SNR is sufficiently high, the interference peak and the noise peak are comparable, while the target peak remains relatively constant at $0$ dB. In such cases, the numerical results of the WD-MAF algorithm can be approximated to those of the MF algorithm.

When the SNR is low, we can analyze the changes in the target peak and the interference peak separately. Firstly, let's consider the target peak. The target peak initially decreases and then increases with decreasing SNR. At an SNR of $-14$ dB, the target peak becomes comparable to the interference peak, which negatively impacts target detection. As the SNR further decreases to $-18$ dB, the target peak approaches the noise peak. This behavior can be attributed to the reduction in the integration space $\mathrm{U}^{(t)}_e$ as the SNR decreases. When the interval length of $\mathrm{U}^{(t)}_e$ becomes smaller than that of $\mathrm{U}^{(t)}_v$, the noise is compensated by $\varsigma^{(t)}$, resulting in a gradual decrease in the target power $\Lambda_s$. As the SNR continues to decrease, the interval length of $\mathrm{U}^{(t)}_e$ tends to zero, and the integrated signal can be considered as noise, leading to $\Lambda_s$ approaching $\Lambda_n$.

Now, let's analyze the interference peak. The interference peak initially increases and then decreases with decreasing SNR. This is because as the SNR decreases, the peak power of the noise gradually approaches that of the interference, making it challenging for the impulse model in the IMM to distinguish between noise and impulse function-induced breakpoints. Consequently, the state estimation performance of the model deteriorates, leading to missed alarms and an increase in $\Lambda_\jmath$. Similarly, as the noise energy contained in $\Lambda_\jmath$ increases with further SNR decrease, $\Lambda_\jmath$ gradually approaches $\Lambda_n$.

Turning to Fig. \ref{fig4}\subref{fig4(b)}, it is observed that as SJR increases, the target peak remains nearly constant at $0$ dB. At high SJRs, the interference peak approaches the noise peak and remains relatively constant. Conversely, at low SJRs, the interference peak initially increases due to the degradation in the model's state estimation performance, as explained earlier.

Based on the aforementioned analysis, it can be concluded that in the simulated scenario, when the SNR is greater than $-8$ dB and the SJR is less than $0$ dB, the numerical results of the WD-AMF algorithm closely align with those of the MF without interference. Furthermore, the sidelobe peaks are lower than $-18$ dB, thereby meeting the detection requirements of the scenario.

\subsection{Parameter sensitivity analysis}

\begin{figure*}[htbp]
\centering
\subfloat[]{
\includegraphics[width=7 cm]{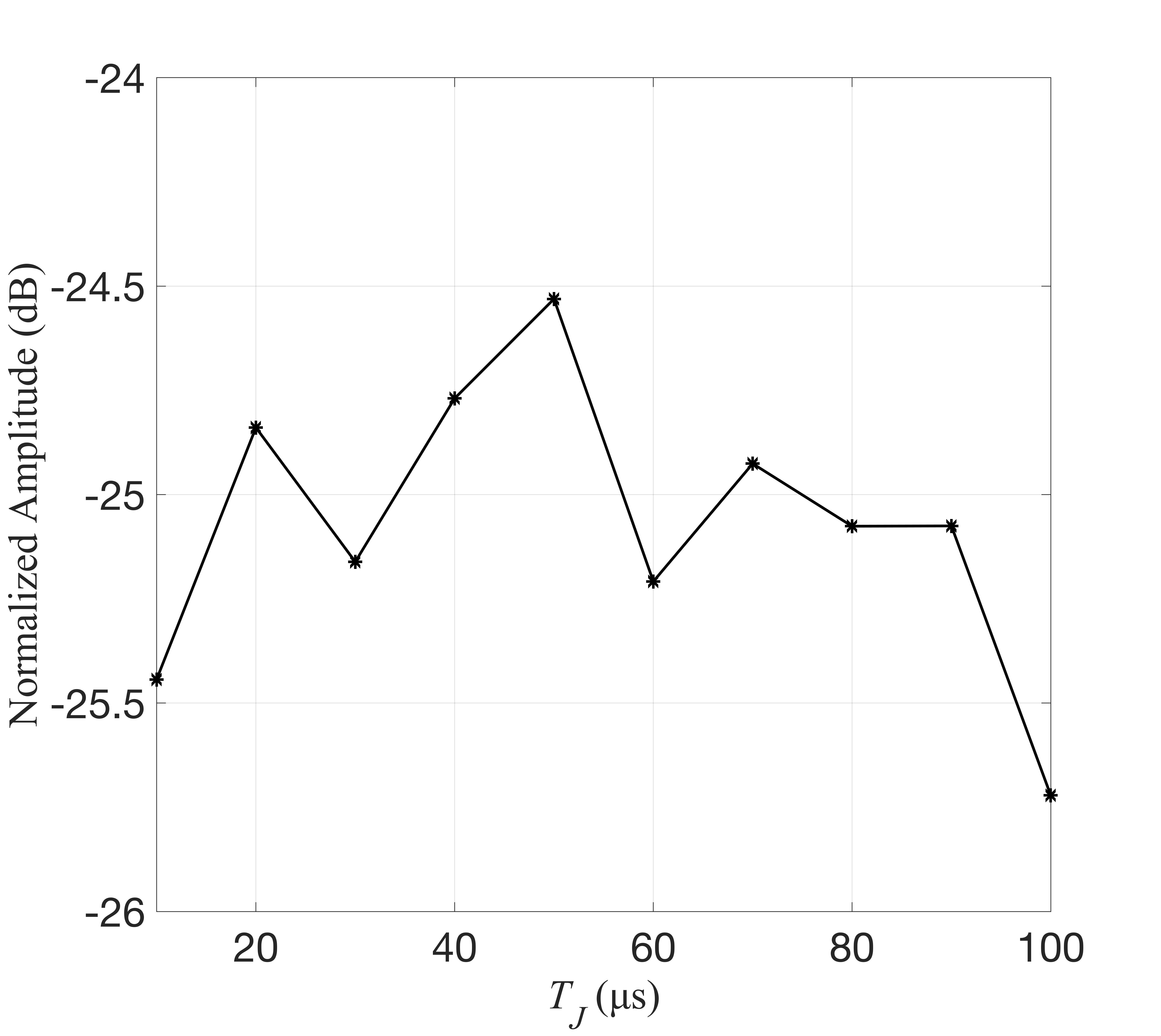}
\label{fig5(a)}}%
\subfloat[]{
\includegraphics[width=7 cm]{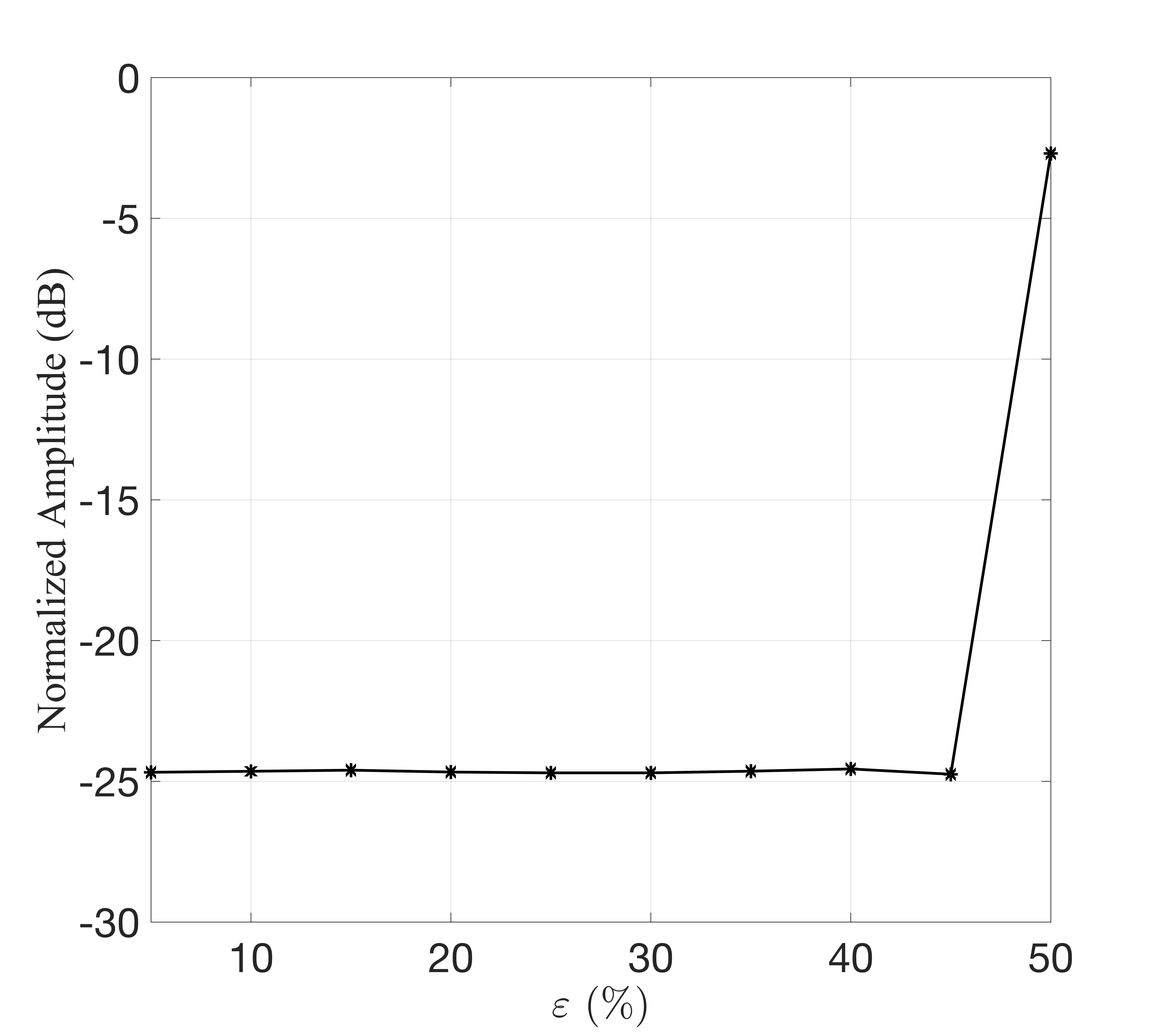}
\label{fig5(b)}}%
\centering
\caption{The curves depicting the variation of peak interference levels under different ISRJ parameters. (a) The curves represent the changes in the levels of the interference peak as a function of $T_J$; (b) The curves represent the changes in the levels of the interference peak as a function of $\varepsilon $.
\label{fig5}}
\end{figure*}

In order to further assess the effectiveness of the proposed method, this section will analyze the method's sensitivity to two crucial parameters of the ISRJ: the sampling repetition period $T_J$ and the sampling duty cycle $\varepsilon$. Experiments will be conducted by varying the $T_J$ and $\varepsilon$ using the simulation scenario parameters specified in Section \ref{S6A}. Fig. \ref{fig5} illustrates the relationship between the average interference peak levels and the varying observational variables.

Fig. \ref{fig5}\subref{fig5(a)} depicts the output peak of the interference signal for different intermittent sampling periods while maintaining a fixed duty cycle of $\varepsilon =20\%$. It is evident that the peak level remains relatively stable around $-25$ dB. This observation indicates that the performance of the proposed WD-AMF is minimally affected by the intermittent sampling period of the ISRJ.

Fig. \ref{fig5}\subref{fig5(b)} illustrates the output peak of the interference signal for various duty cycle conditions while keeping the ISRJ sampling repetition period fixed at $T_{J} = 20$ $\upmu$s. Upon observation, it can be inferred that when the duty cycle $\varepsilon$ is less than $50\%$, the interference peak remains relatively constant, stabilizing at the level of the noise peak. This behavior suggests effective ISRJ suppression during such instances, indicating that the WD-AMF is not significantly influenced by the sampling repetition period. However, when the duty cycle is set to $50\%$, the interference signal experiences a rapid increase, reaching $-3$ dB. This phenomenon signifies the failure of the WD-AMF at a duty cycle of $50\%$. The cause of this failure lies in the fact that at $\varepsilon =50\%$, precisely $A_{\jmath} = E^{(t)}$, rendering (\ref{eq40}) ineffective and resulting in a swift escalation of the interference peak. Consequently, in practical applications, it is advisable to appropriately adjust the adaptive threshold in (\ref{eq40}) to meet the requirements for ISRJ suppression under different duty cycle conditions.

\section{Conclusion}

This paper presents the waveform-Domain adaptive matched filtering (WD-AMF) method as a solution for mitigating interrupted-sampling repeater jamming (ISRJ), aiming to address the limitations of previous matched filtering system-based methods that necessitate urgent ISRJ modeling. By examining the dissimilarities between ISRJ and radar-transmitted waveforms through the cumulative waveform coherence function (CWCF), we identify the primary disparity as the slope difference in the CWCF. We formulate the anti-ISRJ problem by incorporating a CWCF-based objective function and employ the IMM-KF algorithm for state estimation of CWCF. Subsequently, the adaptive weighted function (AWF) in the waveform domain is derived by hypothesis testing, utilizing the target function and estimated state values based on conditional probabilities. The AWF is then utilized to obtain the adaptive filtering term and compensation term. 

Multiple simulations are conducted to demonstrate the effectiveness of the proposed method, showcasing its superior anti-ISRJ performance and adaptability compared to matched filtering system-based methods \cite{9,20}. Parametric sensitivity simulations reveal that WD-AMF exhibits insensitivity to the ISRJ period and duty ratio below $50\%$. 

Nonetheless, the proposed method does possess certain limitations. It assumes the presence of constant modulus constraints or minimal amplitude variations for both the echo signal and the interference signal, which may prove challenging to achieve in the context of wideband signals or a scintillating target. Furthermore, the high Doppler tolerance of LFM waveforms introduces biased estimations during the state estimation phase, hindering accurate estimations of the target function. Hence, it is worthwhile to investigate and discuss potential waveforms characterized by well-defined CWCF.

\section*{Appendix A: CWCF of $s(t)$}

\setcounter{equation}{0}
\renewcommand\theequation{A\arabic{equation}} 
When $-T\leqslant t<0$, and $-\frac{T}{2}\leqslant\rho< \frac{T}{2}+t$,
\begin{equation}
\begin{aligned}
y_{s}^{(t)}(\rho) 
& = \int_{-\frac{T}{2}}^{\rho} e^{j\pi k (t-\mu)^2}\times e^{-j\pi k \mu^2}d \mu
\\
& = c_{1}^{(t)}(\rho) \cdot \operatorname{Sa}\left[\pi k t\left(\rho-\alpha_{s}^{(t)}\right)\right]
\\
& = c^{(t)}(\rho)
\end{aligned}
\label{Aeq2}
\end{equation}
\begin{equation}
\begin{aligned}
c_{1}^{(t)}(\rho)=\left(\rho-\alpha_{s}^{(t)}\right) \cdot e^{j \pi k t\left(t-\rho-\alpha_{s}^{(t)}\right)}
\end{aligned}
\label{Aeq3}
\end{equation} 

When $-T\leqslant t\leqslant0$, and $\frac{T}{2}+t\leqslant\rho\leqslant \frac{T}{2}$,
\begin{equation}
\begin{aligned}
y_s^{(t)}(\rho) 
& = \int_{-\frac{T}{2}}^{\frac{T}{2}+t} e^{j\pi k (t-\mu)^2}\times e^{-j\pi k \mu^2}d \mu
\\
& = c^{(t)}(\beta_s^{(t)}) 
\\
& = c_0^{(t)} 
\end{aligned}
\label{Aeq4}
\end{equation}

When $0<t\leqslant T$, and $-\frac{T}{2}\leqslant\rho< -\frac{T}{2}+t$,
\begin{equation}
\begin{aligned}
y_s^{(t)}(\rho) = 0
\end{aligned}
\label{Aeq6}
\end{equation}

When $0\leqslant t\leqslant T$, and $-\frac{T}{2}+t\leqslant\rho\leqslant \frac{T}{2}$,
\begin{equation}
\begin{aligned}
y_{\mathrm{s}}^{(t)}(\rho) 
& = \int_{-\frac{T}{2}+t}^{\rho} e^{j\pi k (t-\mu)^2 }\times e^{-j\pi k \mu^2}d \mu
\\
& = c_{1}^{(t)}(\rho) \cdot \operatorname{Sa}\left[\pi k t\left(\rho-\alpha_s^{(t)}\right)\right]
\\
& = c^{(t)}(\rho)
\end{aligned}
\label{Aeq7}
\end{equation}

The combination of (\ref{Aeq2})-(\ref{Aeq7}) can be obtained, and the combination of $y_{s}^{(t)}(\rho)$ is expressed as:
\begin{equation}
\begin{array}{l}
y_{\mathrm{s}}^{(t)}(\rho) 
\\ 
=\left\{\begin{array}{cl}
c^{(t)}(\rho) 
& \text { when } \alpha_s^{(t)} \leqslant \rho \leqslant \beta_s^{(t)} \\
c_{0}^{(t)} 
& \text { when } \rho>\beta_s^{(t)} 
\\
0 
& \text { when } \rho<\alpha_s^{(t)}
\end{array}\right.\end{array}
\label{Aeq8}
\end{equation}

\section*{Appendix B: CWCF of $\jmath(t)$}
\setcounter{equation}{0}
\renewcommand\theequation{B\arabic{equation}} 
When $t<-\frac{T+T_\jmath}{2}-nT_{J}$,
\begin{equation}
\begin{aligned}
y_{\jmath_{n}}^{(t)}(\rho) = 0
\end{aligned}
\label{Beq2}
\end{equation}

When $-\frac{T+T_\jmath}{2}-nT_{J}\leqslant t\leqslant-\frac{T-T_\jmath}{2}-nT_{J}$, and $-\frac{T}{2}\leqslant \rho\leqslant\frac{T_\jmath}{2}+nT_{J}+t$
\begin{equation}
\begin{aligned}
y_{\jmath_{n}}^{(t)}(\rho)
& = \int_{-\frac{T}{2}}^{\rho} e^{j\pi k (t-\mu)^2 }\times e^{-j\pi k \mu^2} d \mu
\\
& = b_{n_1}^{(t)}(\rho) \cdot \operatorname{Sa}\left[\pi k t\left(\rho-\alpha_{\jmath_n}^{(t)}\right)\right]
\\
& = b_n^{(t)}(\rho)
\end{aligned}
\label{Beq3}
\end{equation}
\begin{equation}
\begin{aligned}
b_{n_{1}}^{(t)}(\rho)=\left(\rho-\alpha_{\jmath_{n}}^{(t)}\right) \cdot e^{j \pi k t\left(t-\rho-\alpha_{\jmath_{n}}^{(t)}\right)}
\end{aligned}
\label{Beq4}
\end{equation} 

When $-\frac{T+T_\jmath}{2}-nT_{J}\leqslant t\leqslant-\frac{T-T_\jmath}{2}-nT_{J}$, and $\rho>\frac{T_\jmath}{2}+nT_{J}+t$
\begin{equation}
\begin{aligned}
y_{\jmath_{n}}^{(t)}(\rho) 
& = \int_{-\frac{T}{2}}^{\frac{T_\jmath}{2}+nT_{J}+t} A_\jmath e^{j\pi k (t-\mu)^2}\times e^{-j\pi k \mu^2}d \mu
\\
& = b_{n}^{(t)}(\beta_{\jmath_n}^{(t)}) 
\\
& = b_{n_0}^{(t)} 
\end{aligned}
\label{Beq5}
\end{equation}

When $t>-\frac{T-T_\jmath}{2}-nT_{J}$, and $-\frac{T}{2}\leqslant \rho<-\frac{T_\jmath}{2}+nT_{J}+t$
\begin{equation}
\begin{aligned}
y_{\jmath_{n}}^{(t)}(\rho) = 0
\end{aligned}
\label{Beq7}
\end{equation}

When $t>-\frac{T-T_\jmath}{2}-nT_{J}$, and $-\frac{T_\jmath}{2}+nT_{J}+t\leqslant \rho\leqslant\frac{T_\jmath}{2}+nT_{J}+t$
\begin{equation}
\begin{aligned}
y_{\jmath_{n}}^{(t)}(\rho) 
& = \int_{-\frac{T_\jmath}{2}+nT_{J}+t}^{\rho} A_\jmath e^{j\pi k (t-\mu)^2 }\times e^{-j\pi k \mu^2}d \mu
\\
& = b_{n_1}^{(t)}(\rho) \cdot \operatorname{Sa}\left[\pi k t\left(\rho-\alpha_{\jmath_n}^{(t)}\right)\right]
\\
& = b_n^{(t)}(\rho)
\end{aligned}
\label{Beq8}
\end{equation}

When $t>-\frac{T-T_\jmath}{2}-nT_s$, and $\rho>\frac{T_\jmath}{2}+nT_{J}+t$
\begin{equation}
\begin{aligned}
y_{\jmath_{n}}^{(t)}(\rho) 
& = \int_{-\frac{T_\jmath}{2}+nT_J+t}^{\frac{T_\jmath}{2}+nT_J+t} A_\jmath e^{j\pi k (t-\mu)^2 }\times e^{-j\pi k \mu^2}d \mu
\\
& = b_{n_0}^{(t)}
\end{aligned}
\label{Beq9}
\end{equation}

When $t>\frac{T+T_\jmath}{2}-nT_{J}$,
\begin{equation}
\begin{aligned}
y_{\jmath_{n}}^{(t)}(\rho)  = 0
\end{aligned}
\label{Beq10}
\end{equation}

When $\frac{T-T_\jmath}{2}-nT_{J}\leqslant t\leqslant\frac{T+T_\jmath}{2}-nT_{J}$, and $\rho<-\frac{T_\jmath}{2}+nT_{J}+t$,
\begin{equation}
\begin{aligned}
y_{\jmath_{n}}^{(t)}(\rho)  = 0
\end{aligned}
\label{Beq11}
\end{equation}

When $\frac{T-T_\jmath}{2}-nT_{J}\leqslant t\leqslant\frac{T+T_\jmath}{2}-nT_{J}$, and $-\frac{T_\jmath}{2}+nT_{J}+t\leqslant\rho\leqslant \frac{T}{2}$,
\begin{equation}
\begin{aligned}
y_{\jmath_{n}}^{(t)}(\rho) 
& = \int_{-\frac{T_\jmath}{2}+nT_{J}+t}^{\rho} A_\jmath e^{j\pi k (t-\mu)^2 }\times e^{-j\pi k \mu^2}d \mu
\\
& = b_{n_1}^{(t)}(\rho) \cdot \operatorname{Sa}\left[\pi k t+\left(\rho-\alpha_{\jmath_n}^{(t)}\right)\right]
\\
& = b_n^{(t)}(\rho)
\end{aligned}
\label{Beq12}
\end{equation}

When $t<\frac{T-T_\jmath}{2}-nT_{J}$, and $-\frac{T}{2}\leqslant\rho<-\frac{T_\jmath}{2}+nT_{J}+t$,
\begin{equation}
\begin{aligned}
y_{\jmath_{n}}^{(t)}(\rho)  = 0
\end{aligned}
\label{Beq13}
\end{equation}

When $t<\frac{T-T_\jmath}{2}-nT_{J}$, and $-\frac{T_\jmath}{2}+nT_{J}+t\leqslant\rho\leqslant\frac{T_\jmath}{2}+nT_{J}+t$,
\begin{equation}
\begin{aligned}
y_{\jmath_{n}}^{(t)}(\rho) 
& = \int_{-\frac{T_\jmath}{2}+nT_{J}+t}^{\rho} A_\jmath e^{j\pi k (t-\mu)^2 }\times e^{-j\pi k \mu^2}d \mu
\\
& = b_{n_1}^{(t)}(\rho) \cdot \operatorname{Sa}\left[\pi k t\left(\rho-\alpha_{\jmath_n}^{(t)}\right)\right]
\\
& = b_n^{(t)}(\rho)
\end{aligned}
\label{Beq14}
\end{equation}

When $t<\frac{T-T_\jmath}{2}-nT_{J}$, and $\rho>\frac{T_\jmath}{2}+nT_{J}+t$,
\begin{equation}
\begin{aligned}
y_{\jmath_{n}}^{(t)}(\rho) 
& = \int_{-\frac{T_\jmath}{2}+nT_{J}+t}^{\frac{T_\jmath}{2}+nT_{J}+t} A_\jmath e^{j\pi k (t-\mu)^2 }\times e^{-j\pi k \mu^2}d \mu
\\
& = b_{n_0}^{(t)}
\end{aligned}
\label{Beq15}
\end{equation}

The expression for $y_{\jmath_{n}}^{(t)}(\rho)$ can be derived by combining formulas (\ref{Beq2})-(\ref{Beq15}). It can be written as:
\begin{equation}
\begin{array}{l}
y_{\jmath_{n}}^{(t)}(\rho)  
\\ 
=\left\{\begin{array}{cl}
b_n^{(t)}(\rho) 
& \text { when } \alpha_{\jmath_n}^{(t)} \leqslant \rho \leqslant \beta_{\jmath_n}^{(t)} \\ 
b_{n_0}^{(t)} 
& \text { when } \rho>\beta_{\jmath_n}^{(t)} 
\\ 
0 & \text { when } \rho<\alpha_{\jmath_n}^{(t)}
\end{array}\right.\end{array}
\label{Beq16}
\end{equation}












\bibliographystyle{unsrt}
\bibliography{reference}


\begin{IEEEbiographynophoto}{Hanning Su} received the B.Sc degree in electronic engineering from Xidian University in 2018. He is currently working towards the Ph.D. degree in signal and information processing with the National Key Lab of Science and Technology on ATR, National University of Defense Technology. His current research interests  include radar signal processing, target tracking, and radar anti-jamming.
\end{IEEEbiographynophoto}
\begin{IEEEbiographynophoto}{Qinglong Bao} received his B.Sc and Ph.D degrees from the National University of Defense Technology, Changsha, China, in 2003 and 2010, respectively. Currently, he is an Associate Professor with the School of Electronic Science, National University of Defense Technology. His current research interests include radar data acquisition and signal processing.
\end{IEEEbiographynophoto}
\begin{IEEEbiographynophoto}{Jiameng Pan} received the B.E. degree in Zhejiang University in 2013, and the Ph.D. degree in National University of Defense Technology in 2020. He is currently a lecturer with the College of Electronic Science and Technology, National University of Defense Technology. His main research interests include radar signal processing, target tracking, and radar anti-jamming.
\end{IEEEbiographynophoto}
\begin{IEEEbiographynophoto}{Fucheng Guo} received the Ph.D. degree in information and communication engineering from the National University of Defense Technology (NUDT), Changsha, Hunan, China, in 2002.,He is now a Professor in the School of Electronic Science, NUDT. His research interests include source localization, target tracking, and radar/communication signal processing.
\end{IEEEbiographynophoto}
\begin{IEEEbiographynophoto}{Weidong Hu} was born in September 1967. He received the B.S. degree in microwave technology and the M.S. and Ph.D. degrees in communication and electronic system from the National University of Defense Technology, Changsha, China, in 1990, 1994, and 1997, respectively.

He is currently a Full Professor in the ATR Laboratory, National University of Defense Technology, Changsha. His research interests include radar signal and data processing.
\end{IEEEbiographynophoto}

\end{document}